\documentclass[11pt, letterpaper]{article}

\usepackage[letterpaper,left=1in,right=1in,top=1in,bottom=1in]{geometry}
\usepackage{graphicx}
\usepackage{amsmath}
\usepackage{amssymb}
\usepackage{setspace}
\usepackage{epstopdf}
\usepackage{color}
\usepackage{booktabs}
\usepackage{multirow}
\usepackage{cite}
\usepackage{siunitx,etoolbox} % table font bond
\usepackage{arydshln}
\usepackage{subfigure}
\newtheorem{rmk}{Remark}

\title{\Large \bf Reduced-order Koopman modeling and predictive control of nonlinear processes}

\author{
\centerline{\normalsize Xuewen Zhang$^{a}$, Minghao Han$^{a}$, Xunyuan Yin$^{a,}$\thanks{Corresponding author: X. Yin. Tel: (+65) 6316 8746. Email: xunyuan.yin@ntu.edu.sg.}}
\vspace{5mm}\\
\centerline{\small $^{a}$School of Chemistry, Chemical Engineering and Biotechnology, Nanyang Technological University,}\\
\centerline{\small 62 Nanyang Drive, 637459, Singapore}
}
\allowdisplaybreaks

\date{}

\begin{document}

\maketitle
\thispagestyle{empty}
\pagestyle{empty}
\setstretch{1.5}

\begin{abstract}
In this paper, we propose an efficient data-driven predictive control approach for general nonlinear processes based on a reduced-order Koopman operator. A Kalman-based sparse identification of nonlinear dynamics method is employed to select lifting functions for Koopman identification. The selected lifting functions are used to project the original nonlinear state-space into a higher-dimensional linear function space, in which Koopman-based linear models can be constructed for the underlying nonlinear process. To curb the significant increase in the dimensionality of the resulting full-order Koopman models caused by the use of lifting functions, we propose a reduced-order Koopman modeling approach based on proper orthogonal decomposition.
%\textcolor[rgb]{0.00,0.07,1.00}{
%%to provide an efficient means of reducing the high dimensional linear space to a lower-dimensional approximation which reduces computational costs.
%The linearity of the reduced-order Koopman models is well-suited for combination with model predictive control (MPC) applications, but the model approximation errors are important to be considered in the design of the model predictive control.}
A computationally efficient linear robust predictive control scheme is established based on the reduced-order Koopman model. A case study on a benchmark chemical process is conducted to illustrate the effectiveness of the proposed method. Comprehensive comparisons are conducted to demonstrate the advantage of the proposed method.

\end{abstract}

\section{Introduction}

%Large-scale complex chemical processes have been widely used in various industries due to their economic efficiency. Those complex processes face the challenge of combining high dimensionality and nonlinearity, leading to computational or experimental models that are too complex and expensive to control using well-established strategies from modern control theory \cite{arbabi2018data}. To fulfill the operating safety, environmentally sustainable production, and profitability requirements of those complex processes, computationally efficient advanced automatic control systems are needed and should be designed.
Complex industrial processes have been commonly adopted across various industries due to their potential to offer better operating safety, operational efficiency, production consistency, and product quality \cite{daoutidis2018integrating, christofides2013distributed, yin2019subsystem}. To attain these advantages, it is crucial to implement cost-effective advanced process control to appropriately regulate the process operation in real-time. However, the large scales and high nonlinearity of modern industrial processes present challenges in the development and implementation of scalable advanced control schemes \cite{daoutidis2016sustainability, ellis2017economic}.

The development of a successful advanced control system requires a high-fidelity model that is capable of accurately representing the dynamical behavior of the underlying industrial process.
If sufficient physics information is available for establishing nonlinear first-principles models to predict the process dynamics, then nonlinear model predictive control (MPC) represents a widely recognized solution to the control of nonlinear processes. In the context of process control, various nonlinear MPC algorithms and approaches have been proposed for nonlinear processes \cite{henson1998nonlinear, chen1998quasi, cannon2004efficient, maner1996nonlinear, zheng1997computationally}.
However, when the process scale increases with more tightly interconnected physical units being integrated, it becomes overwhelmingly challenging to derive differential equations to accurately characterize their dynamics \cite{jain1997decentralized}. Therefore, the applications of the above-mentioned nonlinear MPC approaches to large and complex industrial processes are still limited. Additionally, even if a first-principles nonlinear model is accessible, the online implementation of these nonlinear MPC algorithms can be demanding. Specifically, the production capacity scales up along with the increase in the number of key states and input variables, which leads to more expensive computation in solving larger-scale (constrained) nonlinear optimization \cite{liu2009distributed}. %{\color{red}[provide a relevant reference that discusses this point, perhaps from a distributed MPC/MHE paper for nonlinear systems]}.
In this work, we aim to address the two limitations of nonlinear system by developing a more efficient approach to control large-scale nonlinear processes. We propose to establish a data-driven linear model to account for nonlinear process dynamics and exploit linear MPC to account for optimal nonlinear process operations.
%\textcolor[rgb]{0.00,0.07,1.00}{Koopman operator was used to form a linear predictor of nonlinear dynamics for designing a linear MPC . To address the model approximation errors during control, robust Koopman MPC have been proposed \cite{wang2022robust, zhang2022robust}.}

The Koopman theory \cite{koopman1931hamiltonian} provides a promising framework for building linear models in a lifted state-space to predict the dynamical behaviors of nonlinear systems/processes \cite{proctor2018generalizing, korda2018linear, das2023koopman, brunton2016koopman, arbabi2018data}. %\cite{proctor2018generalizing, brunton2016koopman, korda2018linear,arbabi2018data}.
An exact Koopman operator can be challenging to be established, and may have infinite dimensions. Therefore, the direct application of Koopman theory to real-world nonlinear processes for simulations and monitoring/decision-making has been impractical. This has motivated the exploration of a finite-dimensional approximation of the exact Koopman operator, and powerful data-driven approximation methods were proposed  \cite{brunton2016koopman, arbabi2018data, schmid2010dynamic}. These approximation methods have greatly facilitated the development of data-driven linear control schemes for general nonlinear systems \cite{korda2018linear,arbabi2018data,wang2022robust}.
In \cite{korda2018linear,arbabi2018data}, Koopman operators were used as linear predictors of nonlinear dynamics, and linear MPC was developed to control nonlinear processes.
In \cite{son2022hybrid}, a hybrid Koopman MPC framework consisting of multiple Koopman models and local controllers was proposed to manage the operation of a pulping process that exhibits differing dynamics.
In \cite{narasingam2019koopman}, the Lyapunov-based MPC framework has been exploited to ensure the closed-loop stability of Koopman-based MPC. Additionally, in \cite{son2022development, narasingam2019koopman}, the Lyapunov function has also been included as one of the lifting functions that map the original nonlinear state-space to a lifted linear state-space for Koopman modeling.
To account for the model mismatch between a Koopman operator and the actual dynamics of the underlying nonlinear system, robust Koopman MPC methods were proposed \cite{wang2022robust, zhang2022robust}.
Offset-free MPC \cite{pannocchia2003disturbance,maeder2009linear} has been integrated with Koopman-based modeling to mitigate the impact of model mismatch arising from data-driven modeling on control performance \cite{son2021application}. In \cite{son2020handling, son2022development}, offset-free Koopman Lyapunov-based MPC methods were proposed, and the stability of the resulting control designs was proven in the presence of plant-model mismatch.

The identification of a Koopman model involves the use of lifting functions to map the original nonlinear state-space into a higher-dimensional linear state-space. The selection of lifting functions can have a substantial impact on both the dimensionality and the accuracy of the resulting Koopman model. Many existing Koopman-based approaches \cite{korda2018linear, narasingam2019koopman}, including the representative extended dynamic mode decomposition \cite{proctor2018generalizing, schmid2010dynamic, narasingam2019koopman}, require a manual selection of the lifting functions based on experience/trial-and-error. However, the manual selection of suitable lifting functions for Koopman models can be very challenging. Additionally, even when appropriate lifting functions can be chosen, the resulting Koopman model will inevitably have a significantly higher dimensionality as compared to the original nonlinear process. While the nonlinear dynamics can be approximated by a linear state-space representation, it will be favorable to maintain the dimensionality of the Koopman model at a manageable level, from an analysis and control perspective. To ensure good prediction performance of the Koopman model and efficient computation of the associated Koopman-based control paradigm, it will be favorable to build an accurate and reasonably-dimensioned Koopman model for optimal control, which entails a systematic approach for selecting appropriate lifting functions, instead of relying on manual selection based on trial-and-error.

%Another challenge associated with Koopman modeling lies in the selection of the lifting functions that relate the original nonlinear state-space to an appropriate lifted linear state-space.
%{\color{blue}The above observations motivate us to propose a new Koopman method that enables automatic selection of appropriate lifting functions to identify an accurate Koopman model, and reducing the dimensionality of the lifted space to facilitate the online implementation of Koopman-based predictive control solutions.}
%The performance of an identified Koopman model may be unsatisfactory if the lifting functions are not appropriately selected.
In the existing literature, to automate the selection of lifting functions for Koopman modeling, machine learning-based algorithms have been proposed \cite{lusch2018deep, schulze2022identification, han2020deep, yeung2019learning, ahmed2023linearizing}. The training of the involved neural networks typically requires large amounts of data. The concept of sparse identification of nonlinear dynamics (SINDy) and Kalman-based recursive algorithms have the potential to be leveraged to automatically select appropriate lifting functions for Koopman modeling. Specifically, SINDy provides a data-based sparse modeling framework, which finds the most active terms for a nonlinear model by solving sparse-promoting optimization \cite{brunton2016discovering, fasel2021sindy, abdullah2023data}. This way, SINDy-based solutions can enhance the model interpretability while avoiding overfitting.
%In addition, a balance between model complexity and model approximation ability can be achieved. The idea of choosing the most active terms for nonlinear modeling can be leveraged to guide the selection of nonlinear lifting functions for Koopman modeling, when sufficient candidate lifting functions are present.
In \cite{wang2022time}, instead of directly solving sparse-promoting optimization problems, a Kalman-based generalized SINDy (called Kalman-GSINDy) method was proposed to recursively identify the most relevant model terms from a library for constructing sparse linear-in-parameter soft sensing models.
The method in \cite{wang2022time} provides a promising alternative pathway to recursively update the most appropriate lifting functions within the Koopman operator framework, instead of determining lifting functions manually based on trial-and-error or prior experience.
%{\color{blue}While the Kalman-GSINDy method has the potential to provide more accurate Koopman models as compared to the Koopman modeling approach based on manual selection of lifting functions, the dimensionality of the resulting Koopman model will inevitably increase as compared to the original nonlinear process. Therefore, for certain processes, in order to prevent the dimensionality of the resulting Koopman models from becoming excessively large, we are further motivated to reduce the dimensionality of the lifted space and the associated linear operator to facilitate efficient online implementation of Koopman-based predictive control solutions.}
In addition to automating the selection of Koopman lifting functions, we also aim to maintain the order of the resulting Koopman model via performing model reduction.
In \cite{lusch2018deep}, deep learning-enabled Koopman modeling using an auto-encoder network was proposed for building reduced-order Koopman models for autonomous systems. Wiener-type Koopman model using deep auto-encoder has been proposed in \cite{schulze2022identification}, where control inputs were explicitly taken into account. These deep learning-based methods may require a large amount of data for training the neural networks, in order to achieve good model accuracy.
%Wiener-type Koopman models for data-driven model reduction has been proposed in \cite{schulze2022identification}. But Koopman models which only have linear dynamics are more universally acceptable compared to Wiener models which consist of both linear and nonlinear dynamics.
Proper orthogonal decomposition (POD) has been widely used for extracting low-dimensional features of high-dimensional systems, which enables a reduction in the dimensionality of the system model being investigated \cite{liang2002proper,rathinam2003new,liu2023state,yin2018state, lumley1967structure,aubry1991hidden, nguyen2020pod, ly2001modeling}. We propose to integrate POD with the Koopman operator to create a reduced-order Koopman modeling framework, which provides linear approximation of nonlinear process dynamics while avoiding a significant increase in dimensionality.

In this work, we propose a data-driven reduced-order Koopman predictive control approach for general nonlinear systems. A Kalman-GSINDy approach is leveraged to select appropriate lifting functions from a comprehensive library. POD is exploited to propose a scalable data-driven Koopman identification approach, which can be used to obtain reduced-order Koopman linear models to approximate the dynamics of the underlying nonlinear process. A linear robust predictive control scheme is proposed based on the reduced-order Koopman model. The effectiveness of the proposed framework is demonstrated based on the application to a reactor-separator process. The superiority of the approach is illustrated in comparison to baselines. Some preliminary results of this work were submitted as a conference paper \cite{zhang2023conferenceversion}. The current paper presents more detailed explanations and discussions of the proposed framework. In addition to simulation results considering the scenario with one single set-point presented in \cite{zhang2023conferenceversion}, the current paper also presents the simulation results considering the scenario with piece-wise constant set-points. Additionally, in the current paper, we include a detailed description of the benchmark process example and present comprehensive comparative results to illustrate the superiority of the proposed method.

\section{Preliminaries}

\subsection{Koopman identification for controlled systems}
Consider a controlled discrete-time nonlinear system of which the dynamics are described by the following nonlinear state-space form:
\begin{equation}\label{paper1:koopman:nonlinear}
x_{k+1} = f \left( x_k, u_k \right)
\end{equation}
where $x_k \in \mathbb{X} \subset \mathbb{R}^n$ is the state vector at sampling instant $k$, $k=1,\ldots, K$; $u_k \in  \mathbb{U} \subset\mathbb{R}^m$ denotes the control input vector at sampling instant $k$; $\mathbb{X}$ and $\mathbb{U}$ are two compact sets.

Based on the Koopman theory for controlled systems, the state is extended to $\chi_k = [x_k^{\mathrm{T}}, u_k^{\mathrm{T}}]^{\mathrm{T}}$. There exists an infinite-dimensional nonlinear lifting mapping $\Psi_{\chi}$, such that the dynamics of the states in the lifted function space are governed by a Koopman operator denoted by $\mathcal K$ as follows \cite{proctor2018generalizing, korda2018linear, williams2016extending}:
\begin{equation}\label{paper1:koopman:linear:1}
	 \Psi_{\chi} ( \chi_{k+1}) = \mathcal{K} \Psi_{\chi} (\chi_k)
\end{equation}
where $\Psi_{\chi}(\chi_k) := [\Psi(x_k)^{\mathrm{T}}, u_k^{\mathrm{T}}]^{\mathrm{T}}$ with $\Psi$ denotes a nonlinear mapping that operates on original state $x$. Instead of attempting to find the precise Koopman operator that can be of infinite dimension, a finite-dimensional nonlinear mapping $\Psi : \mathbb{R}^{n} \rightarrow \mathbb{R}^{N}$ on state $x$ can be considered, and the finite-dimensional Koopman operator $\mathcal{K}$ can be identified on the finite-dimensional space accordingly.
Additionally, finite-dimensional Koopman operator $\mathcal{K}$ can be represented by a block matrix in the following form:
\begin{equation}\label{K:block}
{\mathcal K} = = \left[
    \begin{array}{c;{2pt/2pt}c}
        A&B \\ \hdashline[2pt/2pt]
        *&*
    \end{array}
\right].
\end{equation}\normalsize
We aim to forecast the dynamical behavior of $\Psi$ using the Koopman operator instead of the entire vector $\Psi_{\chi}$. Therefore, it is sufficient to identify matrices $A$ and $B$, which will enable the establishment of the following Koopman based linear model \cite{proctor2018generalizing, arbabi2018data, narasingam2020application}:
%Since the objective of using the Koopman operator is to forecast the dynamical behavior of $\Psi$ but not $u$, we only need to identify $A$ and $B$ instead of identifying the entire operator $\mathcal K$ to establish the following controlled linear Koopman model
\begin{equation}\label{paper1:koopman:linear:2}
	\Psi (x_{k+1}) = A \Psi (x_k) +B u_k.
\end{equation}

\subsection{Proper orthogonal decomposition}\label{paper1:section:POD}
Proper orthogonal decomposition (POD) is a numerical method for obtaining a lower-dimensional approximated representation of a higher-dimensional system/process through data analysis \cite{liang2002proper}. %\cite{chatterjee2000introduction}.
 %POD offers a principled way to analyze multidimensional data which provides an orthonormal basis for representing the used data. The lifting functions resulting from POD are known as basis vectors. %POD can capture the key elements of an infinite-dimensional process with a minimal number of basis vectors in a new state-space.
Based on POD analysis, the original data samples can be expressed in an approximation form by choosing truncated basis vectors from all the basis vectors of the actual samples \cite{rathinam2003new}. %\cite{liang2002proper}

%The objective of POD is to finding a minimal information loss between the original state $v$ and the approximated vector $v^{(r)}$ as given in the following forms:
%\begin{equation}\label{paper1:POD:obj}
%    \begin{split}
%        \min_{\Phi_r} \quad &\mathbb{E} \left\{ \left|\left|v - v^{(r)} \right|\right|^2 \right\}
%      % \mbox{s.t.} \quad &\phi_i^\mathrm{T} \phi_j = \delta_{i,j}, \quad \left( i , j = 1,  2, \cdots, n \right)
%    \end{split}
%\end{equation}
%where $\Phi_r$ is reduced-order projection matrix contains the selected basis vectors.

First, a coordinate transformation is conducted on the original state based on a projection matrix $\Phi \in \mathbb{R}^{n \times n}$ that consists of a set of orthonormal basis vectors, that is, $\Phi := [\phi_1, \phi_2, \ldots, \phi_n]^{\text{T}}$ where $\phi_i$, $i=1,\ldots,n$ is the eigenvector of the covariance matrix based on the data samples. The eigenvalues of the covariance matrix  are sorted in a descending order, such that $\phi_i$ is the eigenvector corresponding to the $i$th largest eigenvalue.
%Let $v \in \mathbb{R}^n$ denotes the original state vector and $\Phi \in \mathbb{R}^{n \times n}$ denotes the projection matrix that consists of a set of orthonormal basis row vectors $\phi_i^{\text{T}}$, $i=1,\ldots,n$ (i.e., $\Phi := [\phi_1^{\text{T}}, \phi_2^{\text{T}}, \ldots, \phi_n^{\text{T}}]^{\text{T}}$).

Next, to reduce the order from $n$ to $r$, the first $r$ rows of $\Phi$ (i.e., the eigenvectors that correspond to the $r$ largest eigenvalues) are selected to form the reduced-order projection matix $\Phi_r \in \mathbb{R}^{r \times n}$, i.e., $\Phi_r= [\phi_1, \phi_2, \ldots, \phi_r]^{\text{T}}$. Based on $\Phi_r$, a reduced-order coordinate transformation can be conducted to obtain a reduced-order approximation of the original state as follows:
\begin{equation}\label{paper1:POD:1}
    q=\Phi_r v
\end{equation}
%POD establishes a vector $\hat{v}$ to approximately represent $v$ based on $q$, defined as follows:
%The relation between reduced-order state $q^{(r)}$ and approximated vector $\hat{z}$ is described as follows:
%then $v$ can be expressed in the form of
% $ \phi_i$, $i=1,\ldots,n$ is a set of orthonormal basis vectors, then $v$ can be expressed in the form of
%\begin{align}
%    \label{paper1:POD:1} v = \Phi q  %= \sum_{i=1}^{n}{q_i \phi_i}
%\end{align}
%where $q \in \mathbb{R}^{n} $ is the state vector in the transformed coordinate.% $\Phi \in \mathbb{R}^{n \times n}$ denotes the collection of the basis vectors that $\Phi := \left[ \phi_1,\ \phi_2,\ \cdots,\ \phi_n \right]$.
%POD establishes a vector $v^{(r)}$ to approximately represent $v$ with a reduced number of basis vectors, defined as follows:
%where $\Phi_r \in \mathbb{R}^{r \times n} $ is reduced-order projection matrix; %of $r$ basis vectors, i.e., $\Phi_r= [\phi_1, \ \phi_2, \ \cdots, \ \phi_r]$.
%$q^{(r)} \in \mathbb{R}^{r}$ is the reduced-order state vector %obtained from $q$ that $q^{(r)}=Mq$ where $M = [I_{r \times r}, 0_{r \times N-r}]$.
where $v \in \mathbb{R}^{n}$ denotes the original state vector; and $q \in \mathbb{R}^{r}$ denotes the reduced-order state in the transformed coordinate.

This way, given a prescribed reduced order $r$, $r < n$, POD establishes the optimal reduced-order projection matrix $\Phi_r$ such that the distance between the original state vector $v$ and its approximation $\hat{v}$ is minimized as follows:
\begin{equation}\label{paper1:POD:obj}
    \begin{split}
        \min_{\Phi_r} \quad &\mathbb{E} \left\{ \left|\left|v - \hat{v} \right|\right|^2 \right\}
      % \mbox{s.t.} \quad &\phi_i^\mathrm{T} \phi_j = \delta_{i,j}, \quad \left( i , j = 1,  2, \cdots, n \right)
    \end{split}
\end{equation}
where $\hat{v}$ is an approximation of $v$ obtained through:
\begin{equation}\label{paper1:POD:3}
\hat{v}  = \Phi_r^+ q= \Phi_r^{\text{T}} q.%, \quad  r < n %= \sum_{i=1}^{r}{q_i \phi_i}
\end{equation}
Additionally, $+$ denotes the matrix pseudo-inverse.

\subsection{Problem formulation}
The objective of this work is twofold: (a) to construct a reduced-order Koopman linear model to describe the dynamics of the nonlinear process in (\ref{paper1:koopman:nonlinear}); (b) to develop a data-driven reduced-order Koopman predictive control scheme for set-point tracking in an efficient manner.

The reduced-order Koopman identification problem can be formulated as the following optimization problem:
% In this work, we aim to construct a reduced-order Koopman linear model and stabilize (\ref{paper1:koopman:nonlinear}) with a data-driven linear robust predictive control scheme.
% % In this work, our objective is (a) to construct a reduced-order Koopman linear model to predict the dynamical behaviors of the underlying nonlinear process in (\ref{paper1:koopman:nonlinear}); (b) to develop a data-driven linear robust predictive control scheme to regulate (\ref{paper1:koopman:nonlinear}) in a more efficient manner based on the reduced-order Koopman model. First, we will construct a linear model in a higher dimensional functional space for nonlinear systems. Then, we reduce the order of the constructed model with a proper collection of lifting functions and corresponding Koopman operators.
% The problem of model identification is formulated in the following form:
\begin{subequations}\label{paper1:problem:linearmodel}
    \begin{align}
       \label{paper1:problem:linearmodel:obj}\min_{A_r, B_r,C, \Phi_r} \sum_{k=1}^{K-1} &\left|\left| x_{k+1} - \hat{x}_{k+1} \right|\right|_2^2 \\
       \label{paper1:problem:linearmodel:cons:1} \mathrm{s.t.} \quad \hat{q}_{k+1} &= A_r \hat{q}_{k} + B_r u_{k} \\
       \label{paper1:problem:linearmodel:cons:2} \hat{x}_k &= C \hat{\Psi}(x_{k}) = C \Phi_r^{\text{T}} \hat{q}_{k}
%       \label{paper1:problem:linearmodel:cons:2} &\hat{\Psi}(x_{k}) = \Phi_r^{\text{T}} \hat{q}_{k}^{(r)} \\
%       \label{paper1:problem:linearmodel:cons:3} &\hat{x}_{k} = C \hat{\Psi}(x_{k})
    \end{align}
\end{subequations}
where $\hat{q}_{k} \in \mathbb{R}^r$ denotes the state vector of the reduced-order Koopman model; $\hat{\Psi}(x_k) \in \mathbb{R}^N$ denotes an approximation of the high-dimensional lifted state vector $\Psi(x_k)$;  $\hat{x}_k \in \mathbb{R}^n$ denotes the predicted state given by the reduced-order Koopman model. The parameters to be identified include system matrices $A_r \in \mathbb{R}^{r \times r}$ and $B_r \in \mathbb{R}^{r \times m}$, state reconstruction matrix $C \in \mathbb{R}^{n \times N}$, and reduce-order projection matrix {$\Phi_r \in \mathbb{R}^{r \times N}$.% need to be identified based on process data.

Based on a reduced-order linear Koopman model in the form of (\ref{paper1:problem:linearmodel}), linear model predictive control will be developed and implemented for the underlying nonlinear process in (\ref{paper1:koopman:nonlinear}).
%we can design the controllers for the identified nonlinear system. In this paper, we want to apply the robust MPC to the linear model to control the original system and test the modeling approximation errors between the original nonlinear system and the reduced-order linear model.

\section{POD-based Koopman modeling using Kalman-GSINDy}
%\section{Architecture of the proposed mechanism}

This section presents the proposed reduced-order Koopman modeling method which integrates the Kalman-generalized sparse identification of nonlinear dynamics (Kalman-GSINDy) algorithm and proper orthogonal decomposition (POD).

\subsection{Architecture of the proposed mechanism}
A graphical illustration of the proposed reduced-order Koopman modeling method is presented in Figure~\ref{paper1:fig:structure}. Kalman-GINDy is leveraged to select appropriate lifting functions that map the original low-dimensional nonlinear state-space of system (\ref{paper1:koopman:nonlinear}) to a higher-dimensional linear state-space. The left block in Figure~\ref{paper1:fig:structure} illustrates the idea of selecting lifting functions based on Kalman-GSINDy. $\mathbf{L}$ is a function library containing all the candidate lifting functions.
%Kalman filter is used to recursively estimate the matrix of coefficients (denoted by $\Omega$) with respect to all the candidate lifting functions.
$\Omega$ is a matrix of coefficients for all the candidate lifting functions with respect to the state variables of the original nonlinear process. Specifically, the $i$th column of $\Omega$, which is denoted by $\xi_i$, contains the coefficients for all the candidate lifting functions with respect to the $i$th variable of the state vector $x$. Kalman filter is used to recursively estimate $\Omega$. The estimate obtained in the final iteration step, denoted by $\hat \Omega_K$, will be used for selecting lifting functions for Koopman modeling.
%The procedure of selecting Koopman lifting functions by Kalman-GSINDy is illustrated in the selection lifting functions section Figure~\ref{paper1:fig:structure}. First, a lifting function library $\mathbf{L}$ that contains all candidate lifting functions is constructed. Next, the Kalman-GSINDy recursively estimates the matrix of coefficients (denoted by $\Omega$) of all the lifting functions. The $i$th column of $\Omega$, which is denoted by $\xi_i$, contains the coefficients for the candidate lifting functions corresponding to the $i$th variable of state vector $x$. The nonlinear mapping consisting of selected lifting functions that will be used for Koopman identification is determined based on the values of the final-step estimate of $\Omega$ (i.e., $\hat{\Omega}_K$).

The right block in Figure~\ref{paper1:fig:structure} illustrates the procedure of reduced-order Koopman modeling based on POD. The higher-dimensional state vector $z$ lifted by the selected lifting functions $\Psi_{KG}$ is projected to a reduced-order state vector $q$ through a projection matrix $\Phi_r$ that is established by conducting POD analysis. Reduced-order Koopman operators $A_r$ and $B_r$ are identified to govern the evolution of $q$ in the reduced-order linear state-space. The prediction of $q$ (denoted by $\hat q$) generated by the reduced-order Koopman model can be used to a predicted state of the original nonlinear process (denoted by $\hat x$).
%The procedure of identifying reduced-order Koopman linear model using POD is presented in the reduced-order Koopman model section in Figure~\ref{paper1:fig:structure}. First, construct the lifted state $z$ and zero-mean lifted state $z'$ using the selected lifting functions. POD is then utilized to obtain the reduced-order state $q$ by finding a reduced-order projection matrix $\Phi_r$. The linearity of the reduced-order state and reduced-order Koopman operator $A_r$, $B_r$ can be identified based on the reduced-order state. At last, the predicted reduced-order state $\hat{q}$ can be reverted back to the original state $\hat{x}$, allowing us to present prediction results based on the reduced-order Koopman model.

\begin{figure*}[t!]
    \centering
    \includegraphics[width=1\textwidth]{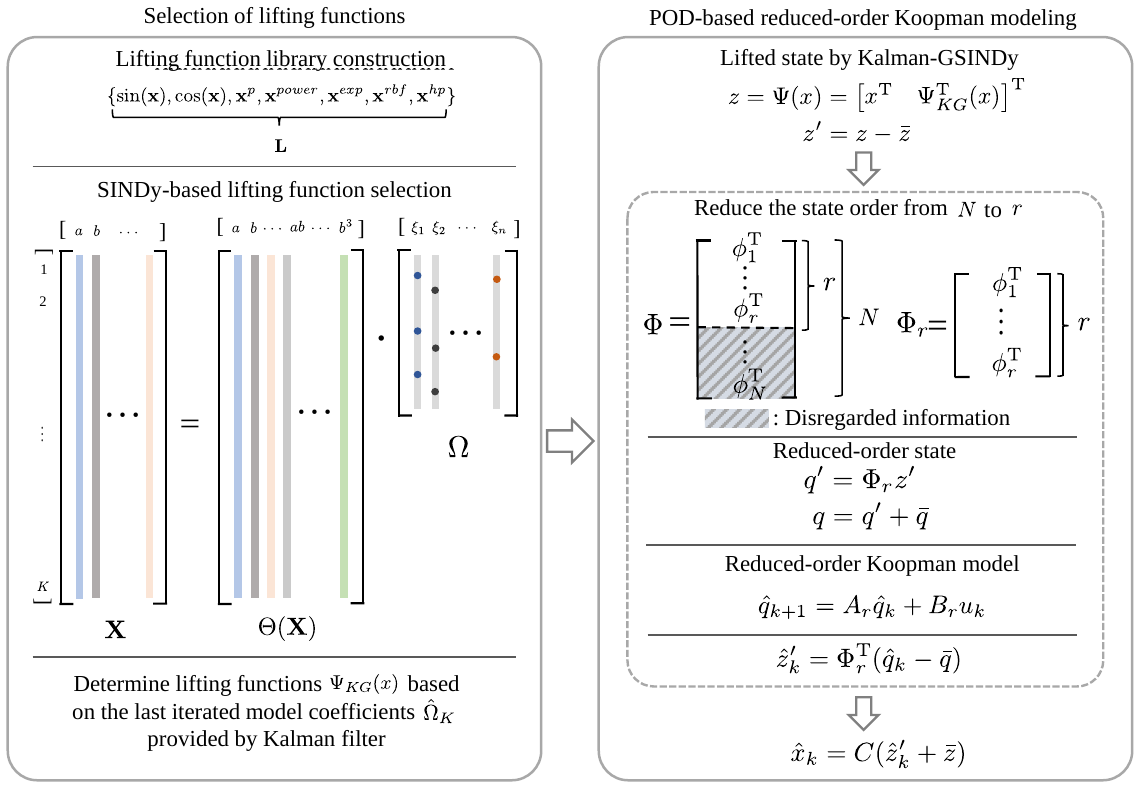}  %{Structure_single.pdf}
   % \captionsetup{font={small}}
    \caption{A graphical illustration of the proposed reduced-order Koopman model for control.}\label{paper1:fig:structure}
\end{figure*}

\subsection{Kalman-GSINDy}

In this subsection, Kalman-GSINDy is exploited to select appropriate lifting functions from an existing rich library as the basis of reduced-order Koopman identification.

In the first step, partially inspired by the existing literature relevant to Koopman modeling and the SINDy algorithm, a lifting function library $\mathbf{L}$ that contains $N_L$ candidate lifting functions is constructed \cite{wang2022time}. With slight abuse of notation, we use row vector $\mathbf{x} \in \mathbb{R}^{1 \times n}$ to denote the transpose of $x$, i.e., $\mathbf{x}=x^{\text{T}}$. A representative lifting function library $\mathbf{L}$ can be created as follows:
%In order to build a universal tool for different systems to find their best suitable lifting functions, first of all, a lifting function library $\Theta(X)$ that contains all potential lifting functions needs to construct. This library has a lot of flexibility and potential lifting functions such as exponential function and polynomial function. %It may include both data-driven terms and potential first-principle terms.
% There are various types of SINDy libraries have been developed including polynomial, trigonometric, Fourier libraries, etc \cite{wang2022time}. The library can be formed in many different forms and can be grouped together to form a new library. Here we build the library $\Theta(X)$ based on the developed SINDy libraries and experience of the Koopman operator. In addition, there are a few broadly applicable functions which are Hermite polynomials, radial basis functions (RBFs), and discontinuous elements \cite{williams2015data}. The lifting function library is given in the following form:
\begin{gather}
    \begin{aligned}
     \mathbf{L} = \{  \mathrm{sin}(\mathbf{x}),  \mathrm{cos}(\mathbf{x}),  \mathbf{x}^{p},  \mathbf{x}^{power},  \mathbf{x}^{exp}, \mathbf{x}^{rbf},  \mathbf{x}^{hp} \}
     %\Theta \left(X \right) = &[ X, \ \mathrm{sin}(X), \ \mathrm{cos}(X), \ X^{poly},  \\
%     &\ X^{power}, \ X^{exp},\ X^{rbf}, \ X^{hp}  ]
\end{aligned}\label{paper1:Kalman SINDy:library}
\end{gather}
%where is a snapshot matrix that containing all data points organized in a temporal order, $K$ is the number of all the samples.
%$X=[x_1, x_2, \cdots, x_n]$ is a state matrix contains $n$ states that $X \in \mathbb{R}^{K \times n}$ and $K$ is the number of all the samples;
where $\mathbf{x}^{p}$ denotes all the terms of possible multiplications of the variables involved in vector $\mathbf{x}$, for example, if $\mathbf{x}=[a, b, c]$, then $\mathbf{x}^p$ includes $ab$, $ac$, $bc$, $abc$;
%the combinations of the variables in the state vector $\mathbf{x}$, e.g., $\mathbf{x}=[a, b,c]$, combinations will be $ab, ac, bc, abc$;
$\mathbf{x}^{power}$ denotes selected power functions; $\mathbf{x}^{exp}$ denotes selected exponential functions; $\mathbf{x}^{rbf}$ and $\mathbf{x}^{hp}$ denote radial basis functions and Hermite polynomials, respectively \cite{williams2016extending, williams2015data}. Note that with slight abuse of notation, each function in (\ref{paper1:Kalman SINDy:library}) conducts element-by-element operations for all the variables involved in $\mathbf{x}$, for example, for state vector $\mathbf{x}=[a,b,c]$, $\text{sin}(\mathbf{x})$ returns $ [\text{sin}(a),\text{sin}(b),\text{sin}(c)]$.
%The notation of functions in (\ref{paper1:Kalman SINDy:library}) has slight abuse that mathematical calculations are element by element, e.g., $\text{sin}(\mathbf{x})$ denotes functions $ \text{sin}(a), \text{sin}(b), \text{sin}(c)  $, assuming state vector $\mathbf{x} = [a, b, c]$ has three variables. %All mathematical calculations are performed element by element.

The next step is to build the dataset matrix $\mathbf{X}$ and lifted dataset matrix $\Theta(\mathbf{X})$. $\mathbf{X} \in \mathbb{R}^{K \times n}$ denotes the matrix of all date samples that $\mathbf{X} =[x_1,\ldots, x_K]^{\text{T}}$. $\Theta (\mathbf{X}) \in \mathbb{R}^{K \times N_L}$ denotes the matrix of the lifted data values. The $k$th row of $\Theta (\mathbf{X})$ is constituted by the values of the candidate lifting functions in $\mathbf{L}$ calculated based on state measurement $\mathbf{x}_k = x_k^{\text{T}}$, $k = 1,\ldots, K$, at time instant $k$ denoted as $\Theta (\mathbf{x}_k)$.
%The next step is to build the lifted dataset matrix $\Theta(\mathbf{X})$ based on the constructed lifting function library $\mathbf{L}$. $\mathbf{X} \in \mathbb{R}^{K \times n}$ denotes the matrix of vector $\mathbf{x}$ that $\mathbf{X}=[\mathbf{x}_1, \ldots, \mathbf{x}_K]^{\text{T}}$. The $k^{\text{th}}$ row of matrix $\Theta(\mathbf{X})$ is constituted by the values of the candidate lifting functions in $\mathbf{L}$ calculated based on state measurement $\mathbf{x}_k = x_k^{\text{T}}$, $k=1,\ldots,K$, at time instant $k$ denoted as $\Theta (\mathbf{x}_k)$.
%is the lifted state vector using library $\mathbf{L}$ at each time instant that denoted as $\Theta(\mathbf{x}_k)$.

The unknown coefficients involved in $\Omega$ are estimated recursively by leveraging the Kalman-GSINDy algorithm \cite{wang2022time}. Specifically, as a new sample $\mathbf{x}_k$ is provided, an updated estimate $\hat{\Omega}_k$ is obtained through a Kalman-like two-step calculation:
\\$Prediction\ step:$
\begin{subequations}\label{paper1: kalman sindy pred}
\begin{align}
\label{paper1: kalman sindy pred: 1}  &\hat{\Omega}_k^- = F_k \hat{\Omega}_{k-1}\\
\label{paper1: kalman sindy pred: 2}  P_k^- = &F_{k-1}P_{k-1}F_{k-1}^\mathrm{T} + Q^*
\end{align}
\end{subequations}
where $\hat{\Omega}_{k-1}$ and $\hat{\Omega}_k^-$ are the posterior and prior estimates of the coefficients for the candidate lifting functions; $P_{k-1}$ and $P_k^-$ denote the posterior and prior estimation error covariance matrices, respectively; $F_k = I$ for $k=1,\ldots,K$;
%$F_k$ is the linear matrix of the candidate lifting function coefficients;
$Q^*$ is the covariance matrix of the process noise.
\\$Correction \ step:$
\begin{subequations}\label{paper1: kalman sindy corr}
\begin{align}
\label{paper1: kalman singdy corr: 1}  K_k = P_k^- \Theta \left( \mathbf{x}_k \right)^\mathrm{T} \Big(\Theta&\left(\mathbf{x}_k \right) P_k^-\Theta \left( \mathbf{x}_k \right)^\mathrm{T} +  R^* \Big)^{-1} \\
%\label{paper1: kalman sindy corr: 2-1}  \Big| \hat{\Omega}_t^- \Big| &< \lambda = 0 \\
\label{paper1: kalman sindy corr: 2-2}  \hat{\mathbf{x}}_k = &\ \Theta \left( \mathbf{x}_k \right) \hat{\Omega}_k^- \\
\label{paper1: kalman sindy corr: 3}  \hat{\Omega}_k = \hat{\Omega}_k^- &+ K_k \left( \mathbf{x}_k - \hat{\mathbf{x}}_k \right) \\
%\label{paper1: kalman sindy corr: 4}  &\hat{\Omega}^-_t = \hat{\Omega}_t \\
\label{paper1: kalman sindy corr: 4}  P_k = ( I - & K_k \Theta \left( \mathbf{x}_k \right) ) P_k^-
\end{align}
\end{subequations}
where $K_k$ is the Kalman gain; $\hat{\mathbf{x}}_k$ represents the estimate of the original state $\mathbf{x}_k$; $\Theta(\mathbf{x}_k)$ is the lifted state value; $R^*$ is the covariance matrix of the measurement noise.

The estimate of the coefficients obtained in the $K$th iteration step (where $K$ is the total number of samples), referred to as $\hat{\Omega}_K$, is treated as the final estimate of $\Omega$, and is used to select appropriate lifting functions from library $\mathbf{L}$ to create $\Psi(x)$. To ensure the estimates of unknown coefficients $\Omega$ will converge and the most appropriate lifting functions will be selected from the library, sufficient data samples of the considered nonlinear process are required, that is, $K$ needs to be sufficiently large.

As shown in Figure~\ref{paper1:fig:structure}, $\Psi(x)$ consists of two components given as follows:
\begin{equation}\label{paper1: kalman sindy: liftfun}
\Psi(x) =
\begin{bmatrix}
          x^{\text{T}} & \Psi_{KG}^{\text{T}}(x)
\end{bmatrix}^{\text{T}}
\end{equation}
where $x$ is the original process state vector and $\Psi_{KG}(x)$ is the vector of nonlinear lifting functions selected from the library based on $\hat{\Omega}_K$. To construct $\Psi_{KG}(x)$, a user-specified positive scalar $\lambda$ is needed. The values of the elements in each row of $\hat {\Omega}_K$ are compared with $\lambda$. If any of the elements in the $i$th row is greater than $\lambda$, then the $i$th candidate lifting function ($i=1,\ldots,N_L$) will be selected and incorporated into $\Psi_{KG}(x)$. Additionally, $z=\Psi(x)$ is determined as the state vector in the lifted state-space.

\begin{rmk}
  The library needs to be constructed in a way such that it contains a diverse range of candidate lifting functions. This ensures the resulting Koopman model, which is established based on the lifting functions selected from the library, can provide good prediction performance.
  The candidate lifting functions in (\ref{paper1:Kalman SINDy:library}) represent some of the commonly used functions.
  We adopt some of the monomials and trigonometric functions from the SINDy libraries considered in \cite{brunton2016discovering, abdullah2023data, wang2022time} to enrich our lifting function library. Exponential functions used in \cite{narasingam2019koopman} for Koopman modeling were also adopted.  To enrich the library with more potential candidate lifting functions, some other functions, such as radial basis functions \cite{williams2016extending} and Hermite polynomials \cite{williams2015data} have been added to the lifting function library.
  The candidate lifting functions included in a library may be adjusted or enriched based on physical knowledge of the specific process being considered, experience of users, and/or trial-and-error analysis.
  It is noted that since only the most appropriate lifting functions will be selected from the candidate lifting functions in the library, further enriching the library with less relevant thus redundant candidate lifting functions will not likely affect the dimensionality of the resulting Koopman model significantly.
   %may adjusted for specific applications based on physical knowledge about specific processes and/or prior experience.  %Other functions may be used to enrich the library.
\end{rmk}

\subsection{POD-based reduced-order Koopman identification}
In this subsection, POD is employed to reduce the dimensionality of the lifted linear state-space, and Koopman operator is identified in the reduced-order state-space to characterize the dynamics of the underlying nonlinear process. %Consequently, the reduced-order Koopman model is formulated to accurately describe the dynamics of the nonlinear process.

After mean removal, zero-mean samples $z'=z-\bar{z}$ (where $\bar{z}$ represents the mean values of the lifted state $z$), can be obtained. Considering (\ref{paper1:POD:1}), the zero-mean reduced-order state $q' \in \mathbb{R}^r$ is given in the following form:
%the following coordinate transformation is conducted:
\begin{equation}\label{paper1:PODKoopman:A}
q' = \Phi_r  z'
\end{equation}
where $\Phi_r \in \mathbb{R}^{r \times N}$ is the reduced-order projection matrix that can be determined following POD analysis in Section~\ref{paper1:section:POD}. The order $r$ of the reduced-order Koopman model is a user-specified parameter, and it needs to be determined carefully based on trial-and-error to balance the trade-off between the model accuracy and the model dimensionality.
Additionally, the mean value of reduced-order state $\bar{q}$ is computed as $\bar{q}=\Phi_r \bar{z}$. %We use $q=q'+\bar{q}$ to denote the reduced-order state with mean value.
Based on (\ref{paper1:koopman:linear:2}), we consider the dynamic of the reduced-order state $q \in \mathbb{R}^{r}$ where $q=q'+\bar{q}$ is governed by reduced-order Koopman operator given as:
%e assume the linearity of the reduced-order state $q \in \mathbb{R}^{r}$ exists given as:
\begin{equation}\label{paper1:PODKoopman:linear}
q_{k+1} = A_r q_k + B_r u_k.
\end{equation}
%where $q_k$ can be computed as $q_k = \Phi_r z_k = q_k'+\bar{q}$.

To construct reduced-order Koopman operator $A_r$ and $B_r$, the following batch least-squares problem is formulated:
\begin{equation}\label{paper1:PODKoopman:obj_mean_1}
%\min_{A_r,B_r} {\sum_{k=1}^{K} \left|\left| \left( q_{k+1}^{(r)'} + \bar{q}^{(r)'} \right) - A_r \left( q_{k}^{(r)} + \bar{q}^{(r)'} \right) - B_r u_k \right|\right|_2^2}
\min_{A_r,B_r} {\sum_{k=1}^{K-1} \left|\left| q_{k+1}  - A_r  q_{k}  - B_r u_k \right|\right|_2^2}.
\end{equation}

Based on the technique adopted in \cite{korda2018linear}, % narasingam2020application},
the analytical solution to (\ref{paper1:PODKoopman:obj_mean_1}) can be obtained:
%The optimization problem (\ref{paper1:PODKoopman:obj_mean_1}) can be solved in the same way of \cite{korda2018linear, narasingam2020application}, and the solution of $A_r$ and $B_r$ can be given in the following form:
\begin{equation}\label{paper1:PODKoopman:sol_mean}
    \relax [A_r, B_r] = \mathbf{q}_{k+1}
      \begin{bmatrix}
         \mathbf{q}_{k}  \\ \mathbf{u}_k
      \end{bmatrix}^{\text{T}}
      \left( \begin{bmatrix}
                \mathbf{q}_{k}  \\ \mathbf{u}_k
             \end{bmatrix}
             \begin{bmatrix}
               \mathbf{q}_{k}  \\\mathbf{u}_k
             \end{bmatrix}^{\text{T}}
      \right)^{+}
\end{equation}
where $\mathbf{q}_{k} \in \mathbb{R}^{r \times K-1}$, $\mathbf{q}_{k+1} \in \mathbb{R}^{r \times K-1}$, and $\mathbf{u}_k \in \mathbb{R}^{m \times K-1}$ are snapshot matrices in the forms of $\mathbf{q}_{k} = [q_{1}, \ldots, q_{K-1}]$, $\mathbf{q}_{k+1} = [q_{2}, \ldots, q_{K} ]$, and $\mathbf{u}_k = [u_1, \ldots, u_{K-1}]$. %In addition, $+$ denotes the pseudo-inverse of a matrix.
%The reduced-order Koopman linear model (\ref{paper1:PODKoopman:prove_linear_2:2}) is identified so that the corresponding $A_r$ and $B_r$ can be solved by (\ref{paper1:PODKoopman:sol_mean}) directly in the reduce-order state-space. Therefore, the optimization step in the Koopman lifted space is omitted.

Based on (\ref{paper1:POD:3}), the relation between the zero-mean reduce-order state $q'$ and the approximated lifted state vector $\hat{z}$ is described as follows:
\begin{equation}\label{paper1:PODKoopman:B}
        \hat{z}  = \Phi_r^{\text{T}} q' + \bar{z}= \Phi_r^{\text{T}}(q-\bar{q})+\bar{z}. %= \hat{z}' + \bar{z} %\Phi_r^{\text{T}} \left(\tilde{q}^{(r)} + \bar{q}^{(r)} \right) = \Phi_r^{\text{T}} q^{(r)}
\end{equation}

Consequently, a reduced-order Koopman model that describes the dynamics of the nonlinear process (\ref{paper1:koopman:nonlinear}) can be formulated:
\begin{subequations}\label{paper1:RKMPC_1}
    \begin{align}
     \label{paper1:RKMPC_1:1}\hat{q}_{k+1} &= A_r \hat{q}_{k}  + B_r u_k  \\
     \label{paper1:RKMPC_1:2}\hat{x}_k &= C \left( \Phi_r^{\text{T}} \big( \hat{q}_{k} - \bar{q} \big) + \bar{z} \right)
    \end{align}
\end{subequations}
where $\hat{q}_{k}$ denotes the predicted reduced-order state of the reduced-order Koopman model;
%$\hat{q}^{(r)'}_k$  denotes the predicted zero-mean reduced-order state, i.e., $\hat{q}^{(r)'}_k = \hat{q}^{(r)}_{k} - \bar{q}^{(r)}$;
%$\hat{z}_k$ denotes the approximated lifted state;
$\hat{x}_k$ denotes the predicted original state of the nonlinear process. In (\ref{paper1:RKMPC_1:2}), the projection matrix $C$ is in the form of $C = [ I_{n \times n}, \ 0_{n \times N-n} ]$. This is because the vector of the lifting functions is determined as (\ref{paper1: kalman sindy: liftfun}), and the first $n$ elements of $z$ are identical to the original nonlinear process states $x$.

\begin{rmk}
    The proposed method is a purely data-driven approach. If any first-principles knowledge is available, it may be favorable to integrate such knowledge with data seamlessly by proposing hybrid modeling approaches \cite{shah2022deep, li2021hybrid, karniadakis2021physics, wu2023physics}. How physical knowledge can be incorporated into the Koopman modeling framework to form physics-enabled Koopman models will be considered in our future work.

\end{rmk}

\section{Robust MPC with reduced-order Koopman operators}

In this section, we propose a robust model predictive control (MPC) scheme based on the reduced-order Koopman in (\ref{paper1:RKMPC_1}) for nonlinear process (\ref{paper1:koopman:nonlinear}).
A block diagram of the proposed reduced-order Koopman robust MPC is presented in Figure~\ref{paper1:fig:RMPC}.

\begin{figure*}[t!]
    \centering
    \includegraphics[width=1\textwidth]{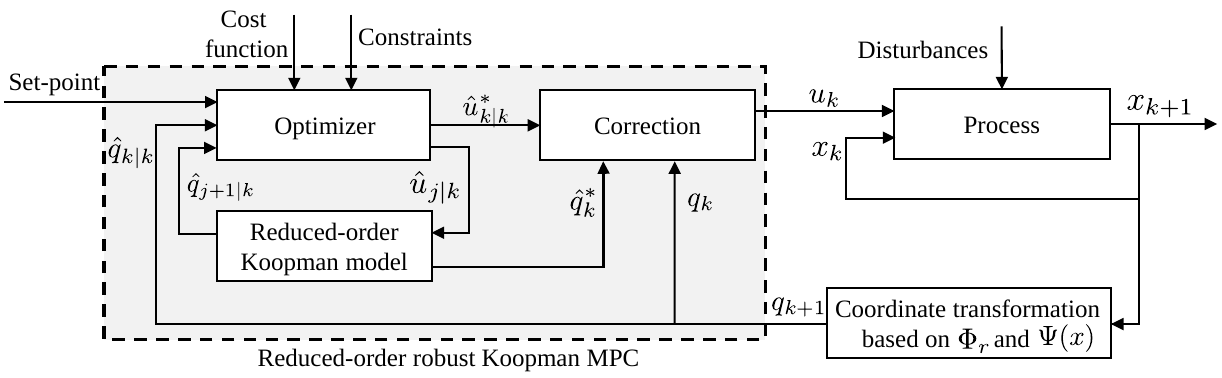}
   % \captionsetup{font={small}}
    \caption{A block diagram of the reduced-order robust Koopman MPC.}\label{paper1:fig:RMPC}
\end{figure*}

First, an MPC design is formulated based on the identified reduced-order Koopman model in the nominal form of (\ref{paper1:RKMPC_1}).
%First, we can formulate the reduced-order Koopman MPC to solve the optimal control action. The reduced-order Koopman model in (\ref{paper1:RKMPC_1}) can be used as the basis of developing the reduced-order Koopman MPC scheme and the nominal system for reduced-order robust Koopman MPC.
The objective function $J$ %for reduced-order Koopman-based predictive control
is formulated as follows:
\begin{equation}\label{paper1:PKMPC:cost:1}
    J = \sum_{j=k}^{k+N_c-1} \ell \left( \hat{q}_{j|k} - q_s , \hat{u}_{j|k} - u_s \right)
\end{equation}
where
\begin{equation}\label{paper1:PKMPC:cost:2}
    \ell \left( \hat{q}_{j|k} - q_s , \hat{u}_{j|k} - u_s \right) = \big|\big| \hat{q}_{j|k} - q_s \big|\big|_Q^2 + \big|\big|  \hat{u}_{j|k} - u_s \big|\big|_R^2
\end{equation}
%\begin{subequations}\label{paper1:PKMPC:cost}
%    \begin{align}
%    \label{paper1:PKMPC:cost:1} J = \sum_{j=k}^{k+N_c-1} &\ell \left( \hat{q}_{j|k} - q_s , \hat{u}_{j|k} - u_s \right)
%    \end{align}
%\text{where}  %\small
%    \begin{align}
%    \label{paper1:PKMPC:cost:2} \ell \left( \hat{q}_{j|k} - q_s , \hat{u}_{j|k} - u_s \right) &= \big|\big| \hat{q}_{j|k} - q_s \big|\big|_Q^2 + \big|\big|  \hat{u}_{j|k} - u_s \big|\big|_R^2
%    \end{align}
%\end{subequations}
\normalsize In (\ref{paper1:PKMPC:cost:2}), $Q \in \mathbb{R}^{r \times r}$ and $R \in \mathbb{R}^{m \times m}$ are positive-definite weighting matrices; $N_c$ is the control horizon; $q_s$ represents the reduced-order state computed based on set-point state $x_s$ that $q_s = \Phi_r \Psi(x_s)$;
%that $x_s = C z_s$ and $ \Phi_r^+ z_s = q_s^{(r)'}+\bar{q}^{(r)'}$ the same as (\ref{paper1:PODKoopman:C}).
$u_s$ is the steady-state reference input corresponding to set-point $x_s$;
%Here $Q \succeq 0$ denotes positive semi-definiteness and $R \succ 0$ denotes positive definiteness.
$\hat{q}_{j|k}$ is a prediction of the reduced-order state for time instant $j$ obtained at time instant $k$;
$\hat{u}_{j|k}$ is the control input for sampling instant $j$ calculated at time instant $k$; $|| x ||^2_Q $ is weighted Euclidean norm of vector $x$, computed as $|| x ||^2_Q:= x^{\text{T}}Qx$.
%In addition, $|| x ||^2_Q := x^{\text{T}}Qx$ denotes the 2-norm weighted by $Q$. Finally, the MPC problem can be summarized as follows:\small
Based on the objective function in (\ref{paper1:PKMPC:cost:1}), a reduced-order MPC design is formulated as follows:
\begin{subequations}\label{paper1:RKMPC:optim}
%\begin{align}
% \label{paper1:RKMPC:optim:1} \min_{\hat{u}_{k|k}, \ldots, \hat{u}_{k+N_c-1|k}} &J
%\end{align}
\begin{align}
    \label{paper1:RKMPC:optim:1} &\min_{\hat{u}_{k|k}, \ldots, \hat{u}_{k+N_c-1|k}} J \\
   \text{s.t.} \quad &\hat{q}_{j+1|k} = A_r  \hat{q}_{j|k}  + B_r \hat{u}_{j|k}, \\
   &\hat{q}_{k|k} = \Phi_r \Psi(x_k),\\ %&\hat{q}_{k|k} = q_{k}=\Phi_r \Psi(x_k),\\
   &C \left( \Phi_r^{\text{T}} \big (\hat{q}_{j|k} - \bar{q} \big) + \bar{z}\right) \in \mathbb{X},  \\%\quad j = k,\ldots, k+N_c \\%\quad j = k,\ldots, k+N_c \\
   &\hat{u}_{j|k} \in \mathbb{U} , \quad j = k,\ldots, k+N_c-1.%\\
\end{align}
\end{subequations}
\normalsize
In (\ref{paper1:RKMPC:optim}), $ \hat{u}_{k|k}, \ldots, \hat{u}_{k+N_c-1|k} $ denotes the control input sequence; $q_{k}$ is the reduced-order state obtained from the state measurement $x_{k}$ at time $k$. % following (\ref{paper1:PODKoopman:4}).
%$\mathbb{X}$, and $\mathbb{U}$ are the non-empty, convex, compact sets of the nonlinear process states $x$ and manipulated inputs $u$.
The optimal solution to problem (\ref{paper1:RKMPC:optim}) at time instant $k \in \mathbb{N}$ is represented by:
\begin{equation}\label{paper1:RKMPC_solu_1}
    %\hat{\mathbf{u}}^{*}_k = \{ \hat{u}_{k|k}^{*}, \hat{u}_{k+1|k}^{*}, \ldots, \hat{u}_{k+N_c|k}^* \}
     \hat{u}_{k|k}^{*}, \hat{u}_{k+1|k}^{*}, \ldots, \hat{u}_{k+N_c-1|k}^*.
\end{equation}
The first control action $\hat{u}_{k|k}^*$ in the optimal control sequence in (\ref{paper1:RKMPC_solu_1}) is applied to the reduced-order Koopman model in (\ref{paper1:RKMPC_1:1}) to generate predicted state $\hat{q}_{k+1}^{*}$, which will be further used to compute the optimal control action that needs to be applied to the nonlinear process in (\ref{paper1:koopman:nonlinear}).
%And the initial reduced-order state $\hat{q}_{k,0}^{(r)*}$ can be obtained.
Then, to account for the model mismatch between the reduced-order Koopman model and the actual dynamics of the nonlinear process (\ref{paper1:koopman:nonlinear}), a stochastic version of the reduced-order Koopman linear model is presented as follows:
\begin{subequations}\label{paper1:RKMPC_2}
    \begin{align}
     \label{paper1:RKMPC_2:1}q_{k+1} &= A_r q_{k}   + B_r u_k  + w_k\\
     %\label{paper1:RKMPC_2:2}z_k &= \Phi_r^{\text{T}} \big( q^{(r)}_k - \bar{q}_k^{(r)} \big) + \bar{z} + v_k\\
%     \label{paper1:RKMPC_2:3}x_k &= C z_k
     \label{paper1:RKMPC_2:2}x_k &= C \left( \Phi_r^{\text{T}} \big( q_k - \bar{q} \big) + \bar{z} \right) + v_k
%     \label{paper1:RKMPC_2:1}q_{k+1}^{(r)} &= A_r \left( q_{k}^{(r)}  + \bar{q}^{(r)'} \right) + B_r u_k - \bar{q}^{(r)'} + w_k\\
%     \label{paper1:RKMPC_2:2}z_k &= \Phi_r \left( q_{k}^{(r)} + \bar{q}^{(r)'} \right) + v_k\\
%     \label{paper1:RKMPC_2:3}x_k &= C z_k
    \end{align}
\end{subequations}
where $w_k$ and $v_k$ account for time-varying model uncertainties. We assume that the model uncertainties $w_k$ and $v_k$ are bounded such that $w_k \in \mathbb{W}, \ v_k \in \mathbb{V},\ \forall k \in \mathbb{N}$ where $\mathbb{W}$ and $\mathbb{V}$ are compact sets. Note that this type of assumption on the boundedness of the system uncertainties was also made in relevant work on robust MPC, see, e.g., \cite{wang2022robust, mayne2005robust}. % ,zhang2022robust}.

The error between the actual and nominal reduced-order state, i.e., (\ref{paper1:RKMPC_2:1}) and (\ref{paper1:RKMPC_1:1}), is computed as:
%given by the reduced-order Koopman model in (\ref{paper1:RKMPC_2:1}) and predicted reduced-order state given by (\ref{paper1:RKMPC_1:1}) is presents in the following form:
%For the reduced-order Koopman model in (\ref{paper1:RKMPC_1:1}), the error between the actual reduced-order state (\ref{paper1:RKMPC_2:1}) and its prediction given by the reduced-order Koopman model is with the following form:
\begin{equation}\label{paper1:RKMPC_error}
    e_{k,q} = q_k - \hat{q}_k^{*}.
\end{equation}

To account for the model uncertainties of the reduced-order Koopman model, following the robust MPC algorithm in \cite{wang2022robust}, we introduce a state-feedback control and form the actual controller output $u_k$ as:
%the optimal robust MPC control action $u_k$ at time step $k$ is determined based on the solution to (\ref{paper1:RKMPC:optim}) as follows:
%To address the model error, we assume there exists a local control law for the reduced-order Koopman linear model. The optimal control action $u_k$ at time step $k$ is given by \cite{wang2022robust}
%$Local \ Control \ Law$: For the POD based Koopman linear Model, there exists a local control law
\begin{equation}\label{paper1:RKMPC:lcl}
u_k = \hat{u}_{k|k}^* + K_r e_{k,q} %U_k = \hat{U}_k + K \left( Q_{k,l} - \hat{Q}_{k,l} \right)
\end{equation}
%where $\hat{u}_{k|k}^*$ is the manipulated input given by MPC at time step $k$; $\hat{q}_{k}^{(r)*}$ is the predicted reduced-order state given by MPC at time step $k-1$; $q_k^{(r)}$ is the reduced-order state obtained from $x_k$ the closed-loop state measurement at time step $k$.
where $K_r \in \mathbb{R}^{m \times r}$ is the state-feedback gain matrix. %The optimal control action $u_k$ is applied to the nonlinear process in (\ref{paper1:koopman:nonlinear}) to regulate the process operation.

Based on (\ref{paper1:RKMPC:lcl}), the error dynamics of $e_{k,q}$ can be expressed as $e_{k+1, q} = A_K e_{k, q}+w_k$, where $A_K := A_r + B_r K_r$. If $K_r$ is determined appropriately such that $A_K$ is Schur stable, then the error dynamics are stable, and the tracking error asymptotically converges to zero. %For the sake of brevity, the detailed stability analysis of the closed-loop system is referred to \cite{mayne2005robust}.

\section{Application to a reactor-separator process}
In this section, we apply the proposed method to a benchmark chemical reaction process via simulations to illustrate the proposed framework.
%effectiveness and superiority of the reduced-order Koopman identification and data-driven control based on the reduced-order Koopman linear model.

%A detailed description of this process and the process model can be found in \cite{liu2009distributed}. In this set of simulations, the liquid hold-up in each of the three vessels is expected to remain at a constant level. The control objective is to track a desired set-point by manipulating the heat inputs $Q_i$, $i=1,2,3$, to the three vessels.
\subsection{Process description and process operation objective}

The benchmark chemical process involves two continuous stirred reactors (CSTRs) and one flash tank separator; a schematic diagram of this process is presented in Figure~\ref{paper1:fig:2cstr}. In this process, two irreversible reactions take place simultaneously: the conversion of reactant $\mathbf{A}$ into desired product $\mathbf{B}$, and the conversion of a portion of $\mathbf{B}$ into side product $\mathbf{C}$. The three vessels of this process are interconnected with each other through mass and energy flows. Specifically, a fresh feed flow carrying pure reactant $\mathbf{A}$ is introduced into the first CSTR (CSTR 1) with volumetric flow rate $F_{10}$ and temperature $T_{10}$. The output stream from CSTR 1 flows into the second CSTR (CSTR 2); another feed stream carrying pure $\mathbf{A}$ enters CSTR 2 at flow rate $F_{20}$ and temperature $T_{20}$. The two chemical reactions $\mathbf{A}\rightarrow \mathbf{B}$ and $\mathbf{B}\rightarrow \mathbf{C}$ take place in the two CSTRs. The effluent from CSTR 2 is directed to the separator at flow rate $F_2$ and temperature $T_2$ \cite{liu2009distributed, zhang2013distributed}.

\begin{figure}[tb]
    \centering
    \includegraphics[width=0.7\textwidth]{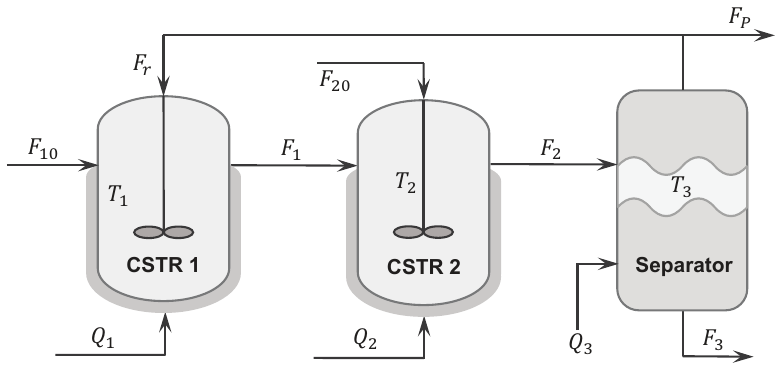}
    %\captionsetup{font={small}}
    \caption{Reactor-separator process.}
    \label{paper1:fig:2cstr}
\end{figure}

The separator provides a recycle stream that re-enters the first vessel at flow rate $F_r$ and temperature $T_3$. Each of the three vessels is equipped with a heating jacket, which is used to add heat to or remove heat from the corresponding vessel at a variable heat input rate $Q_i$, $i = 1, 2,  3$. The state variables of this nonlinear include the mass fractions of materials $\mathbf{A}$ and $\mathbf{B}$, denoted by $X_{\mathbf{A}i}$ and $X_{\mathbf{B}i}$, and the temperatures $T_i$, $i=1,2,3$, in the three vessels. Nine ordinary different equations have been established based on the materials and energy balances to characterize the dynamical behaviors of the process states, which are shown as follows \cite{liu2009distributed, zhang2013distributed,li2023iterative}:
%This model, together with a more detailed description of this chemical process can be found in \cite{zhang2013distributed}. %\cite{liu2009distributed}.

\begin{subequations}\label{paper1: cstr:equ}
\begin{align}
%&~~\frac{dV_1}{dt} = F_{f1}+F_R-F_1\\[0.3em]
& \displaystyle{\frac{dx_{\mathbf{A}1}}{dt}} = \frac{F_{10}}{V_1}(x_{\mathbf{A}10} - x_{\mathbf{A}1}) + \frac{F_r}{V_1}(x_{\mathbf{A}r} - x_{\mathbf{A}1}) - k_1 e^{\frac{-E_1}{rT_1 }}x_{\mathbf{A}1}  \\[0.3em]
& \displaystyle{\frac{dx_{\mathbf{B}1}}{dt}} = \frac{F_{10}}{V_1}(x_{\mathbf{B}10} - x_{\mathbf{B}1}) +  \frac{F_r}{V_1}(x_{\mathbf{B}r} - x_{\mathbf{B}1})+  k_1 e^{\frac{-E_1}{rT_1 }}x_{\mathbf{A}1} - k_2 e^{\frac{-E_2}{rT_1 }}x_{\mathbf{B}1}  \\[0.3em]
& \displaystyle{~~\frac{dT_{1}}{dt}} = \frac{F_{10}}{V_1}(T_{10} - T_{1}) + \frac{F_r}{V_1}(T_{3} - T_{1}) - \frac{\Delta H_1}{ c_p}k_{1}e^{\frac{-E_1}{rT_1}}x_{\mathbf{A}1} - \frac{\Delta H_2}{ c_p}k_{2}e^{\frac{-E_2}{rT_1}}x_{\mathbf{B}1} +  \frac{Q_1}{\rho c_pV_1} \\[0.3em]
%&~~\frac{dV_2}{dt} = F_{f2}+F_1-F_2\\[0.3em]
& \displaystyle{\frac{dx_{\mathbf{A}2}}{dt}} = \frac{F_{1}}{V_2}(x_{\mathbf{A}1} - x_{\mathbf{A}2}) + \frac{F_{20}}{V_2}(x_{\mathbf{A}20} - x_{\mathbf{A}2}) - k_1 e^{\frac{-E_1}{rT_2 }}x_{\mathbf{A}2}  \\[0.3em]
& \displaystyle{\frac{dx_{\mathbf{B}2}}{dt}} = \frac{F_{1}}{V_2}(x_{\mathbf{B}1} - x_{\mathbf{B}2}) + \frac{F_{20}}{V_2}(x_{\mathbf{B}20} - x_{\mathbf{B}2})+  k_1 e^{\frac{-E_1}{rT_2 }}x_{\mathbf{A}2} - k_2 e^{\frac{-E_2}{rT_2}}x_{\mathbf{B}2}  \\[0.3em]
& \displaystyle{~~\frac{dT_{2}}{dt}} = \frac{F_{1}}{V_2}(T_{1} - T_{2}) + \frac{F_{20}}{V_2}(T_{20} - T_{2}) - \frac{\Delta H_1}{ c_p}k_{1}e^{\frac{-E_1}{rT_2}}x_{\mathbf{A}2} - \frac{\Delta H_2}{ c_p}k_{2}e^{\frac{-E_2}{rT_2}}x_{\mathbf{B}2} +  \frac{Q_2}{\rho c_pV_2} \\[0.3em]
%&~~\frac{dV_3}{dt} = F_{2}-F_P-F_R-F_3\\[0.3em]
& \displaystyle{\frac{dx_{\mathbf{A}3}}{dt}} = \frac{F_{2}}{V_3}(x_{\mathbf{A}2} - x_{\mathbf{A}3}) - \frac{(F_r+F_p)}{V_3}(x_{\mathbf{A}r} - x_{\mathbf{A}3})  \\[0.3em]
& \displaystyle{\frac{dx_{\mathbf{B}3}}{dt}} = \frac{F_{2}}{V_3}(x_{\mathbf{B}2} - x_{\mathbf{B}3}) - \frac{(F_r+F_p)}{V_3}(x_{\mathbf{B}r} - x_{\mathbf{B}3}) \\[0.3em]
& \displaystyle{~~\frac{dT_{3}}{dt}} = \frac{F_{2}}{V_3}(T_{2} - T_{3}) + \frac{Q_3}{\rho c_pV_3}+\frac{(F_{r}+F_{p})}{\rho c_{p}V_{3}}(x_{\mathbf{A}r}\Delta H_{\text{vap1}}+x_{\mathbf{B}r}\Delta H_{\text{vap2}}+x_{\mathbf{C}r}\Delta H_{\text{vap3}} )
%+ \frac{(F_R+F_p)}{\rho c_pV_3} (x_{AR}\Delta H_{vap1} + x_{BR}\Delta H_{vap2} + x_{CR}\Delta H_{vap3})
\end{align}
\end{subequations}
%\textcolor[rgb]{0.00,0.07,1.00}{where $x_{\mathbf{C}3}$ is the mass fraction of $\mathbf{C}$ in the separator; $x_{\mathbf{A}10}$, $x_{\mathbf{B}10}$ are the percent compositions of $\mathbf{A}$ and $\mathbf{B}$ in the feed flows; $X_{\mathbf{A}r}$, $X_{\mathbf{B}r}$, $X_{\mathbf{C}r}$ are the mass fractions of $\mathbf{A}$, $\mathbf{B}$, $\mathbf{C}$ in the recycle; $F_1$, $F_2$ are the effluent flow rates from vessels 1, 2; $F_r$, $F_p$ are the flow rates of the recycle and purge; $V_1$, $V_2$, $V_3$ are the volumes of vessels 1, 2, 3; $E_1$, $E_2$ are the activation energy for reactions 1, 2; $k_1$, $k_2$ are the pre-exponential values for reactions 1,2; $\Delta H_1$, $\Delta H_2$ are heats of reaction for reactions 1, 2; $c_p$ is the heat capacity; $r$ is the gas constant; $\rho$ is solution density.}
%The model of the separator vessel was derived based on the assumption that there is a negligible amount of reaction taking place in the separator. It has also been assumed that the relative volatility for each of the components remains constant within the operating temperature range. The assumptions allow calculating the mass fractions in the liquid portion of the vessel.
In addition, the algebraic equations that describe the relationship between the composition of the overhead stream and the composition of the liquid holdup in the separator are given as follows:
\begin{equation}\label{paper1: cstr: alg equ}
\begin{aligned}
x_{\mathbf{A}r} &= \frac{\alpha _\mathbf{A} x_{\mathbf{A}3}}{\alpha _\mathbf{A} x_{\mathbf{A}3} + \alpha _\mathbf{B} x_{\mathbf{B}3} + \alpha _\mathbf{C} x_{\mathbf{C}3} } \\[0.3em]
x_{\mathbf{B}r} &= \frac{\alpha _\mathbf{B} x_{\mathbf{B}3}}{\alpha _\mathbf{A} x_{\mathbf{A}3} + \alpha _\mathbf{B} x_{\mathbf{B}3} + \alpha _\mathbf{C} x_{\mathbf{C}3} } \\[0.3em]
x_{\mathbf{C}r} &= \frac{\alpha _\mathbf{C} x_{\mathbf{C}3}}{\alpha _\mathbf{A} x_{\mathbf{A}3} + \alpha _\mathbf{B} x_{\mathbf{B}3} + \alpha _\mathbf{C} x_{\mathbf{C}3} } %\\[0.3em]
%x_{\mathbf{C}3} &= 1-x_{A3}-x_{B3}
\end{aligned}
\end{equation}
The definitions for the process variables are given in Table~\ref{paper1:table:process variables}. A detailed description of this chemical process and the values of the model parameters can be found in \cite{liu2009distributed, zhang2013distributed}. %In this set of simulations, the liquid hold-up in each of the three vessels is expected to remain at a constant level. The control objective is to track a desired set-point by manipulating the heat inputs $Q_i$, $i=1,2,3$, to the three vessels.

The objective is: 1) to establish a Koopman linear model of a moderate dimensionality to describe the dynamics of the nonlinear chemical process; 2) to drive the process operation to desired set-points by regulating the heat inputs $Q_i$, $i=1,2,3$, applied to the three vessels.

\begin{table*}[t!]
  \renewcommand\arraystretch{1.2}
  \caption{Process variables.}\label{paper1:table:process variables}
  \centering%\small
    \begin{tabular}{m{5cm} l }
      \hline
      $x_{\mathbf{A}1}$, $x_{\mathbf{A}2}$, $x_{\mathbf{A}3}$ & Mass fractions of $\mathbf{A}$ in CSTR 1, CSTR 2, separator \\
      $x_{\mathbf{B}1}$, $x_{\mathbf{B}2}$, $x_{\mathbf{B}3}$ & Mass fractions of $\mathbf{B}$ in CSTR 1, CSTR 2, separator \\
      $x_{\mathbf{C}3}$, $x_{\mathbf{C}3}$, $x_{\mathbf{C}3}$ & Mass fractions of $\mathbf{C}$ in CSTR 1, CSTR 2, separator \\
      $x_{\mathbf{A}r}$, $x_{\mathbf{B}r}$, $x_{\mathbf{C}r}$ & Mass fractions of $\mathbf{A}$, $\mathbf{B}$, $\mathbf{C}$ in the recycle stream \\
      $x_{\mathbf{A}10}$, $x_{\mathbf{B}10}$                  & Mass fractions of $\mathbf{A}$, $\mathbf{B}$ in the feed flows \\
      $T_1$, $T_2$, $T_3$                                     & Temperatures in CSTR 1, CSTR 2, separator \\
      $T_{10}$, $T_{20}$                                      & Feed stream temperatures to CSTR 1, CSTR 2, separator \\
      $F_1$, $F_2$                                            & Effluent flow rates from CSTR 1, CSTR 2 \\
      $F_{10}$, $F_{20}$                                      & Steady-state feed stream flow rates to CSTR 1, CSTR 2 \\
      $F_r$, $F_p$                                            & Flow rates of the recycle and purge streams \\
      $V_1$, $V_2$, $V_3$                                     & Volumes of CSTR 1, CSTR 2, separator \\
      $E_1$, $E_2$                                            & Activation energy for reactions $\mathbf{A} \rightarrow \mathbf{B}$, $\mathbf{B} \rightarrow \mathbf{C}$ \\
      $k_1$, $k_2$                                            & Pre-exponential values for reactions $\mathbf{A} \rightarrow \mathbf{B}$, $\mathbf{B} \rightarrow \mathbf{C}$ \\
      $\Delta H_1$, $\Delta H_2$                              & Heats of reaction for reactions $\mathbf{A} \rightarrow \mathbf{B}$, $\mathbf{B} \rightarrow \mathbf{C}$ \\
      $\Delta H_{vap1}$, $\Delta H_{vap2}$, $\Delta H_{vap3}$ & Evaporating enthalpies for $\mathbf{A}$, $\mathbf{B}$, $\mathbf{C}$ \\
      $\alpha_{\mathbf{A}}$, $\alpha_{\mathbf{B}}$, $\alpha_{\mathbf{C}}$ & Relative volatilities of $\mathbf{A}$, $\mathbf{B}$, $\mathbf{C}$ \\
      $Q_1$, $Q_2$, $Q_3$                                     & Heat inputs into CSTR 1, CSTR 2, separator \\
      $c_p$, $r$, $\rho$                                      & Heat capacity, gas constant and solution density \\
      \hline
    \end{tabular}
\end{table*}

\subsection{Simulation setting}
%We consider an open-loop stable steady-state as the set-point, which is shown in Table~\ref{paper1:table:states:x}. This set-point $x_{s}$ corresponds to constant heat inputs $u_{s}$ where $Q_{1,s} = 2.87 \times 10^6~\text{KJ/h}$, $Q_{2,s} = 1.00 \times 10^6~\text{KJ/h}$, and $Q_{3,s} = 2.87 \times 10^6~\text{KJ/h}$. $x_0$ in Table~\ref{paper1:table:states:x} is used as the initial condition for closed-loop simulation based on the proposed control method.

The liquid hold-up in each of the three vessels remains at a constant level during the process operation. We consider two open-loop stable steady-state levels as the set-points to illustrate the proposed framework. The two set-points $x_{s1}$ and $x_{s2}$ are shown in Table~\ref{paper1:table:states:x}. Each of the set-points is corresponding to one steady-state level of heat inputs, which is presented in Table~\ref{paper1:table:states:u}. %In addition, $x_0$ in Table~\ref{paper1:table:states:x} is used as the initial condition for closed-loop simulations based on the proposed control method.

\begin{table*}[tb!]
  \renewcommand\arraystretch{1.25}
  \caption{The initial state $x_0$ and the two set-points $x_{s1}$, $x_{s2}$.}\label{paper1:table:states:x}
  \centering%\small
    \begin{tabular}{ c c c c c c c c c c }
      \toprule
      States &  $x_{A1}$ & $x_{B1}$ &  $T_{1} \ (\text{K})$ & $x_{A2}$ & $x_{B2}$ & $T_{2} \ (\text{K})$ & $x_{A3}$ & $x_{B3}$ & $T_{3} \ (\text{K})$  \\
      \midrule
      $x_{0}$ &
        0.1155 & 0.6235 & 497.3 &
        0.1367 & 0.6053 & 489.8 &
        0.0396 & 0.5504 & 491.8 \\
      $x_{s1}$ &
        0.1921 & 0.6753 & 476.8 &
        0.2117 & 0.6561 & 468.5 &
        0.0721 & 0.6896 & 471.5 \\
      $x_{s2}$ &
        0.1336 & 0.6475 & 491.4 &
        0.1547 & 0.6284 & 483.9 &
        0.0469 & 0.5956 & 486.0 \\

      \bottomrule
    \end{tabular}
\end{table*}

\begin{table}[t!]
  \renewcommand\arraystretch{1.25}
  \caption{Heat inputs $u_{s1}$, $u_{s2}$ corresponding to the set-points $x_{s1}$, $x_{s2}$.}\label{paper1:table:states:u}
  \centering%\small
    \begin{tabular}{ c c c c c c c c c c }
      \toprule
      Heat inputs & $Q_1~(\text{kJ/h})$ & $Q_2~(\text{kJ/h})$ & $Q_3~(\text{kJ/h})$  \\
      \midrule
      $u_{s1}$ &
      $2.87 \times 10^6$ & $1.00 \times 10^6$ & $2.87 \times 10^6$ \\
      $u_{s2}$ &
      $2.94 \times 10^6$ & $1.14 \times 10^6$ & $2.95 \times 10^6$ \\
      \bottomrule
    \end{tabular}
\end{table}

%\begin{table}[t!]
%  \renewcommand\arraystretch{1.25}
%  \caption{The upper bounds and lower bounds of heat inputs $u_{max}$, $u_{min}$.}\label{paper1:table:states:ubound}
%  \centering%\small
%    \begin{tabular}{ c c c c c c c c c c }
%      \toprule
%      Bounds of heat inputs & $Q_1~(\text{kJ/h})$ & $Q_2~(\text{kJ/h})$ & $Q_3~(\text{kJ/h})$  \\
%      \midrule
%      % after \\: \hline or \cline{col1-col2} \cline{col3-col4} ...
%      $u_{max}$ &
%      $3.49 \times 10^6$ & $1.20 \times 10^6$ & $3.49 \times 10^6$ \\
%      $u_{min}$ &
%      $2.465 \times 10^6$ & $0.85 \times 10^6$ & $2.465 \times 10^6$ \\
%      \bottomrule
%    \end{tabular}
%\end{table}

\begin{table}[t!]
  \renewcommand\arraystretch{1.25}
  \caption{The upper bounds and lower bounds of heat inputs $u_{max}$, $u_{min}$.}\label{paper1:table:states:ubound}
  \centering%\small
    \begin{tabular}{ c c c c c c c c c c }
      \toprule
      Bounds of heat inputs & $Q_1~(\text{kJ/h})$ & $Q_2~(\text{kJ/h})$ & $Q_3~(\text{kJ/h})$  \\
      \midrule
      $u_{max}$ &
      $2.976 \times 10^6$ & $1.026 \times 10^6$ & $2.976 \times 10^6$ \\
      $u_{min}$ &
      $2.85 \times 10^6$ & $0.98 \times 10^6$ & $2.85 \times 10^6$ \\
      \bottomrule
    \end{tabular}
\end{table}

\begin{figure}[tb]
    \centering
    \includegraphics[width=0.9\textwidth]{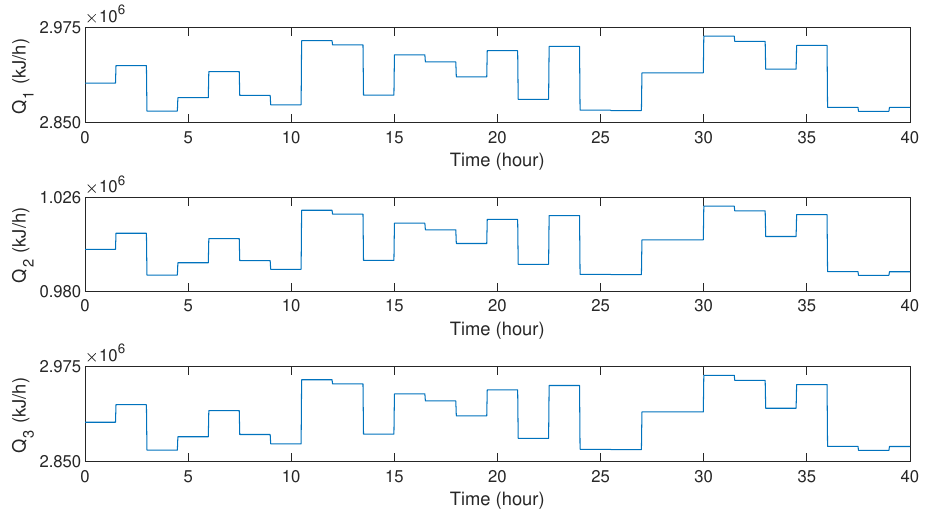}
    %\captionsetup{font={small}}
    \caption{Trajectories of the heat inputs generated for modeling.}
    \label{paper1:fig:utrajectory}
\end{figure}

The state measurements are sampled asynchronously at time instant $k$ with a sampling period of 0.025 hours.
%such that \textcolor[rgb]{1.00,0.00,0.00}{$ \{t_{k \geq 0} \}$ such that $t_k = t_0 + k \Delta$} with sampling period $\Delta = 0.025 \ \text{hr}$.
%with $t_0 = 0$ the initial time, $\Delta$ a fixed sampling time of controller and sensor which is chosen to be $\Delta = 0.025 \ \text{h} = 1.5 \ \text{min}$, and $k$ positive integers.
%To obtain the Koopman lifting functions,
First, open-loop simulations are conducted to generate data for Koopman modeling. The heat inputs applied to the three vessels are generated randomly in a uniform distribution within the prescribed ranges and are varied after every 1.5 hours, which are shown in Figure~\ref{paper1:fig:utrajectory}.
%The maximum heat input $u_{max}$ is set as $Q_{1,max} = 3.49 \times 10^6~\text{kJ/h}$, $Q_{2,max} = 1.20\times 10^6~\text{kJ/h}$, and $Q_{3,max} = 3.49 \times 10^6~\text{kJ/h}$. The minimum heat input $u_{min}$ is set as $Q_{1,min} = 2.465\times 10^6~\text{kJ/h}$, $Q_{2,min} = 0.85\times 10^6~\text{kJ/h}$, and $Q_{3,min} = 2.465 \times 10^6~\text{kJ/h}$.
The upper bounds and lower bounds on the open-loop heat inputs are shown in Table~\ref{paper1:table:states:ubound}.
Bounded random disturbances are added to the process. The disturbances added to the dynamics of the concentrations are generated as normally distributed values with zero mean and a standard deviation of 1, and are then made bounded within the interval $[-0.5, 0.5]$. The disturbances added to the dynamics of the temperature are generated as normally distributed values with a zero mean and a standard deviation of 10, and are then made bounded within $[-5, 5]$.

The construction of the library $\mathbf{L}$ is guided by (\ref{paper1:Kalman SINDy:library}). Specifically, the library comprises power functions and exponential functions including $\mathbf{x}^{-0.5}, \mathbf{x}^{0.5}, \mathbf{x}^2, \mathbf{x}^{2.5}, \mathbf{x}^3, \ldots, \mathbf{x}^5, (1+0.25\mathbf{x}^2)^{-0.5}, (1+0.25\mathbf{x}^2)^{0.5}, e^{\mathbf{x}},e^{-0.2\mathbf{x}^2}$, and $\mathbf{x}^{p}$, $\mathbf{x}^{rbf}$, and $\mathbf{x}^{hp}$ functions.% are also considered and incorporated into the library.% Additionally, all mathematical calculations are performed element by element.

\subsection{Reduced-order Koopman modeling}
By implementing the Kalman-GSINDy algorithm, 16 lifting functions are selected from the library containing 610 lifting functions, which include the original state variables of the nonlinear process (\ref{paper1:koopman:nonlinear}), and 7 lifting functions $\Psi_{KG}(x) = [ x_{(1)} x_{(3)} x_{(4)}, x_{(1)} x_{(3)} x_{(4)} x_{(6)}, x_{(3)} x_{(5)} x_{(6)} x_{(7)}, x_{(5)} x_{(7)} x_{(8)} x_{(9)}, \\  x_{(1)} x_{(2)} x_{(4)} x_{(7)} x_{(8)},  x_{(2)} x_{(3)} x_{(6)} x_{(7)} x_{(9)},$ $\Pi_{i=1}^{9}x_{(i)}]^{\text{T}}$, where $x_{(i)}$ denotes the $i$th variable of state vector $x$. Then, after applying the proposed reduced-order Koopman modeling approach, the order is reduced from 16 to 8 and a reduced-order Koopman linear model with 8 latent state variables is constructed.

\begin{figure*}[htb]
    \centering
    \includegraphics[width=1\textwidth]{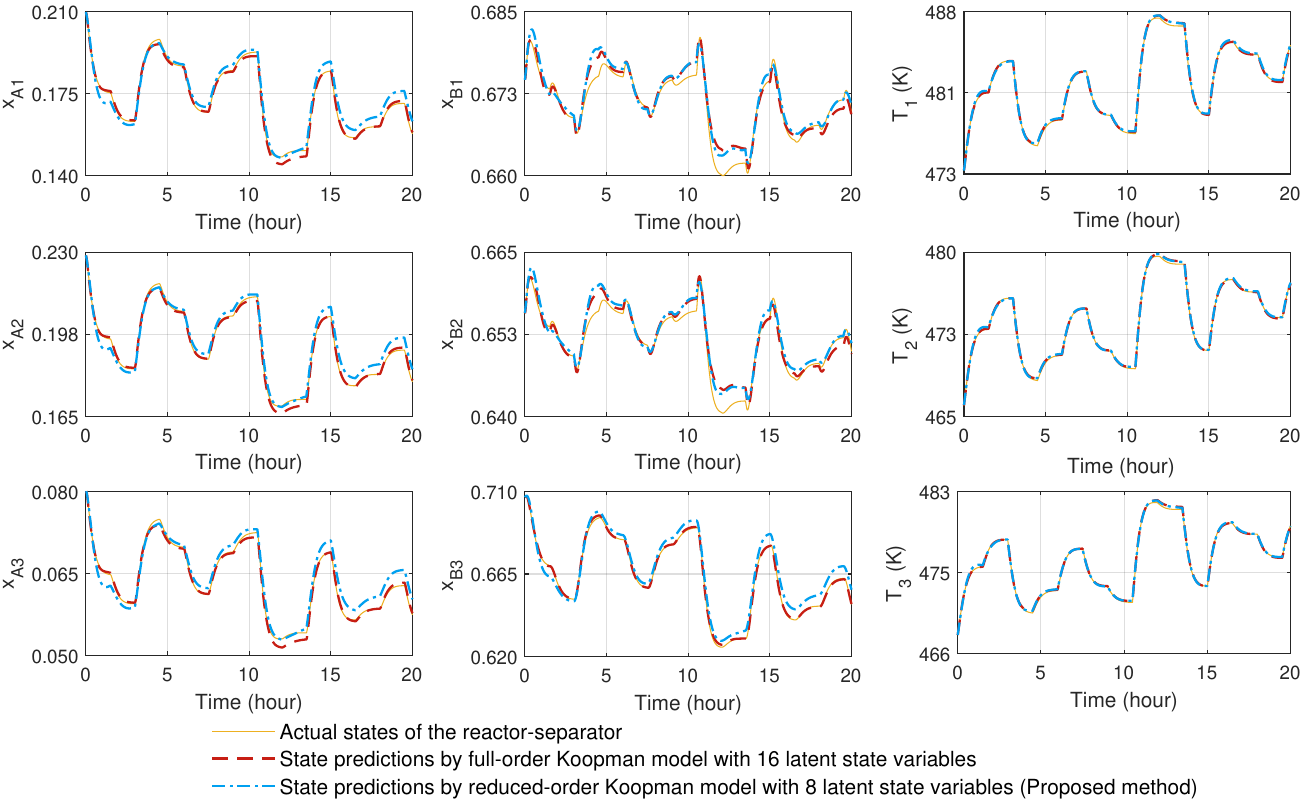}
    %\captionsetup{font={small}}
    \caption{Prediction trajectories under the full-order Koopman and reduced-order Koopman model. The order of the full-order Koopman state is $N=16$, and the order of the reduced-order Koopman state is $r=8$.}
    \label{paper1:fig:modeling}
\end{figure*}

\begin{figure*}[h!]
    \centering
    \subfigure[State trajectories under the full-order KMPC, reduced-order KMPC, and reduced-order RKMPC.]{
    \label{paper1:fig:2cstr:1setpoint_x}
    \includegraphics[width=0.98\textwidth]{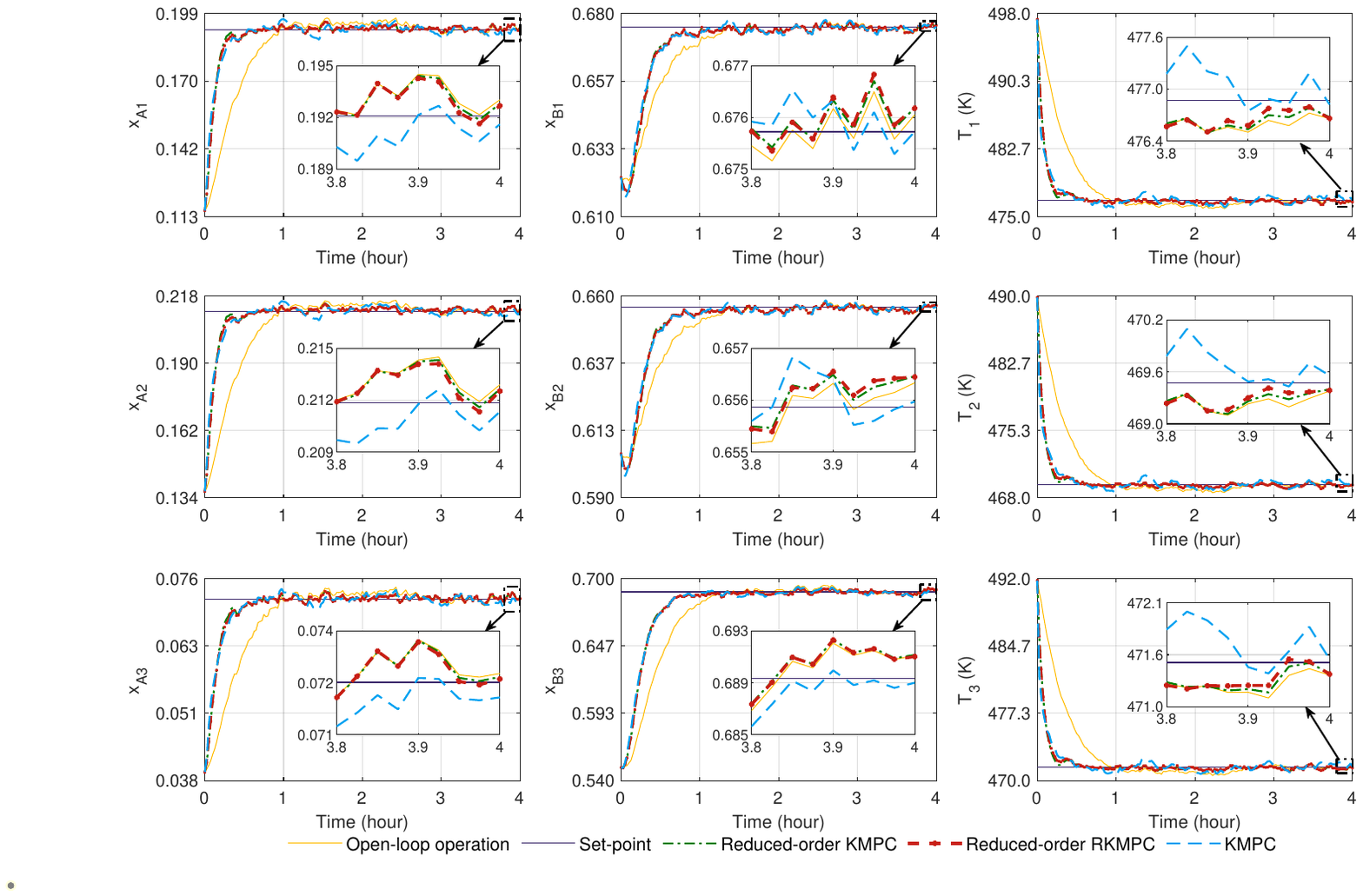}
    %\captionsetup{font={small}}
    }

    \subfigure[Manipulated input trajectories under the full-order KMPC, reduced-order KMPC, and reduced-order RKMPC.]{
    \label{paper1:fig:2cstr:1setpoint_u}
    \includegraphics[width=0.98\textwidth]{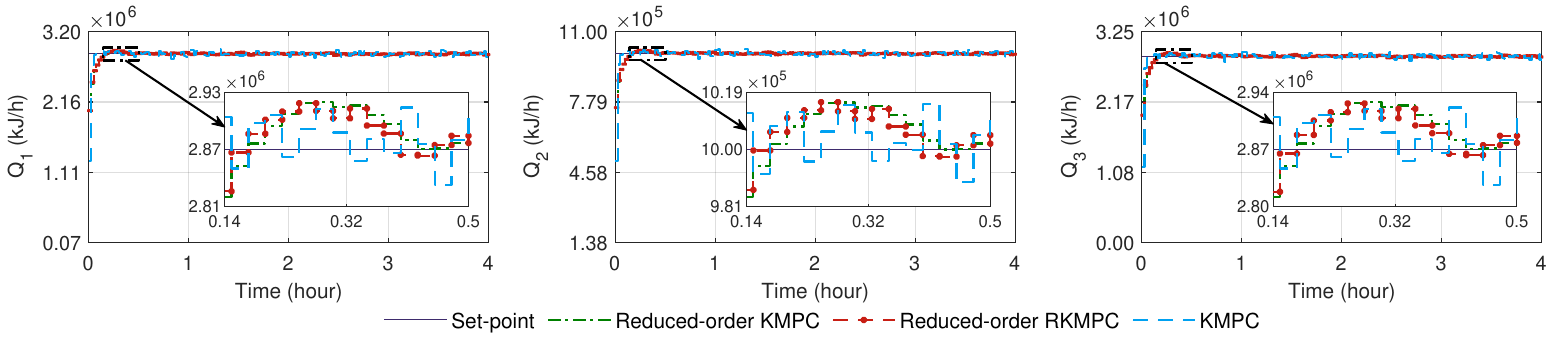}
    }
    \caption{State and manipulated input trajectories with one single set-point.}
    \label{paper1:fig:2cstr:1setpoint}
\end{figure*}

%\begin{figure*}[h!]
%    \centering
%    \includegraphics[width=0.9\textwidth]{1setpoint_x.pdf}
%    \caption{State trajectories under the full-order KMPC, reduced-order KMPC, and reduced-order RKMPC.}
%    \label{paper1:fig:2cstr:1setpoint_x}
%\end{figure*}
%
%\begin{figure*}[h!]
%    \centering
%    \includegraphics[width=0.9\textwidth]{1setpoint_u.pdf}
%    \caption{Manipulated input trajectories under the full-order KMPC, reduced-order KMPC, and reduced-order RKMPC.}
%    \label{paper1:fig:2cstr:1setpoint_u}
%\end{figure*}

\begin{figure*}[h!]
    \centering
    \subfigure[State trajectories under the full-order KMPC, reduced-order KMPC, and reduced-order RKMPC.]{
    \label{paper1:fig:2cstr:2setpoint_x}
    \includegraphics[width=0.98\textwidth]{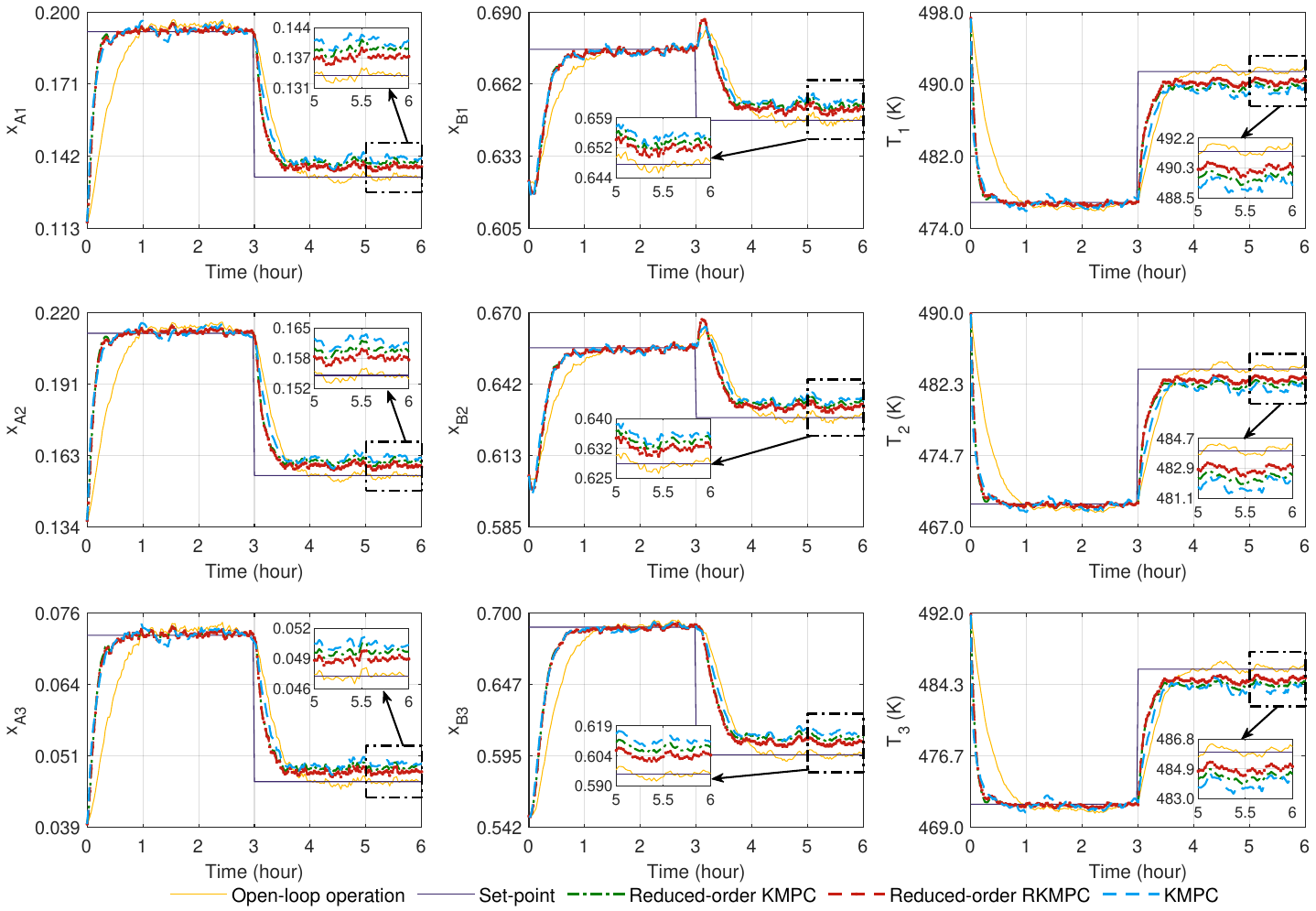}
    %\captionsetup{font={small}}
    }

    \subfigure[Manipulated input trajectories under the full-order KMPC, reduced-order KMPC, and reduced-order RKMPC.]{
    \label{paper1:fig:2cstr:2setpoint_u}
    \includegraphics[width=0.98\textwidth]{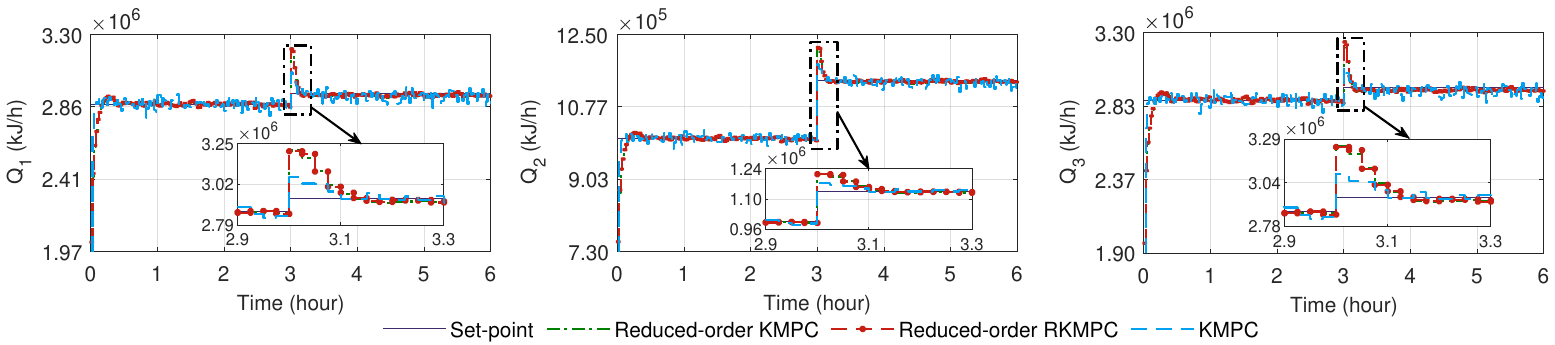}
    }
    \caption{State and manipulated input trajectories with a change in set-point.}
    \label{paper1:fig:2cstr:2setpoint}
\end{figure*}

The open-loop predictions of the state variables of the reactor-separator process given by the proposed method, and the actual state trajectories are shown in Figure~\ref{paper1:fig:modeling}. The predictions (in blue dash-dotted lines) of the reduced-order Koopman model can track the trends of actual state trajectories. For comparison, Figure~\ref{paper1:fig:modeling} also shows the state predictions of a full-order Koopman model with 16 latent state variables. This model is constructed based on the same lifting functions adopted in reduced-order Koopman modeling, yet it does not perform model order reduction. This full-order Koopman model provides slightly better prediction accuracy than the proposed method. However, the usage of a full-order Koopman model will lead to a significant increase in the computational complexity for the associated MPC, which will be discussed in Section~\ref{paper1:section:cost}.

\subsection{Control performance}\normalsize

In this section, we evaluate and compare three control schemes: the proposed reduced-order robust Koopman MPC (reduced-order RKMPC); reduced-order Koopman MPC (reduced-order KMPC); full-order Koopman MPC (full-order KMPC) without model reduction. The control objective is to track the set-points $x_{s1}$ and $x_{s2}$ given in Table~\ref{paper1:table:states:x}, and maintain the process operation around the respective level. Two different case scenarios are considered to evaluate the control performance of different control schemes, that is, the case scenario with one single set-point $x_{s1}$ to track, and the case scenario with a change in the set-point from $x_{s1}$ to $x_{s2}$ after three-hour process operation.
%The first case considers a constant set-point $x_{s1}$ for the controllers to track. In the second case, the set-point will be changed from $x_{s1}$ to $x_{s2}$, allowing for observing the performance of the controllers during the transition period between two set-point.

The control horizons for the three control designs are set as $N_c = 15$. For the full-order KMPC, the weighting matrices are $Q_{full} = \mathrm{diag}([12.34,  7.21, 8.44,  10.45,  11.23, 9.22,  10.23,   6.44,  8.54, 12.45, \\ 9.32,  13.43,   12.43,  9.54,  11.32, 8.23]) \in \mathbb{R}^{16 \times 16}$ and $R_{full} = \mathrm{diag}([2.8,  3.3,  2.6]) \in \mathbb{R}^{3 \times 3}$, where $diag(\cdot)$ denotes a diagonal matrix. The weighting matrices of the reduced-order KMPC and RKMPC models are $Q_{reduced}=\mathrm{diag}([37.34,  38.21,  40.44,   49.45,48.23,  37.22, 38.23, 31.44 ])  \in \mathbb{R}^{8 \times 8}$ and $R_{reduced}= \mathrm{diag}([2.8,  3.3, 2.6]) \in \mathbb{R}^{3 \times 3}$.

First, we consider the case when only $x_{s1}$ is considered as the reference to track. Figure~\ref{paper1:fig:2cstr:1setpoint_x} presents the closed-loop state trajectories generated by the three controllers, while Figure~\ref{paper1:fig:2cstr:1setpoint_u} presents the corresponding control input trajectories given by the three controllers. All the three control designs are able to steer the process states towards the desired set-point, and maintain the operation level close to the set-point.

Figure~\ref{paper1:fig:2cstr:2setpoint} presents the closed-loop state trajectories under the other case scenario with set-point change during process operation. Figure~\ref{paper1:fig:2cstr:2setpoint_x} presents the closed-loop state trajectories generated by the three controllers, and Figure~\ref{paper1:fig:2cstr:2setpoint_u} presents the trajectories of the control inputs by the three controllers. All the three controllers are capable of tracking the piecewise-constant set-points. The proposed reduced-order RKMPC outperforms the other two designs in the sense that it can bring the closed-loop states closer to the reference values.
%As shown in Figure~\ref{paper1:fig:2cstr:2setpoint_x}, all the three control designs can effectively stabilize the process states after the set-point transition. The proposed reduced-order Koopman-based robust MPC approach demonstrated the best operation performance.

\begin{figure*}[t!]
    \centering
    \subfigure[3 RMSEs of different operating periods under the scenario with one single set-point.]{
    \label{paper1:fig:RMSE:case:1sp}
    \includegraphics[width=1\textwidth]{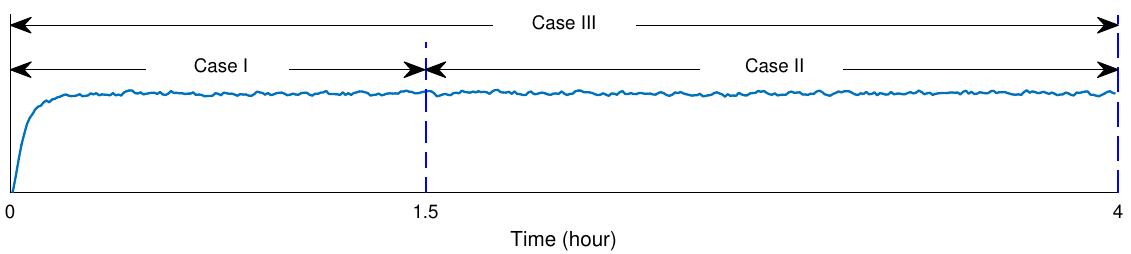}
    %\captionsetup{font={small}}
    }

    \subfigure[3 RMSEs of different operating periods under the scenario with a change in the set-point.]{
    \label{paper1:fig:RMSE:case:2sp}
    \includegraphics[width=1\textwidth]{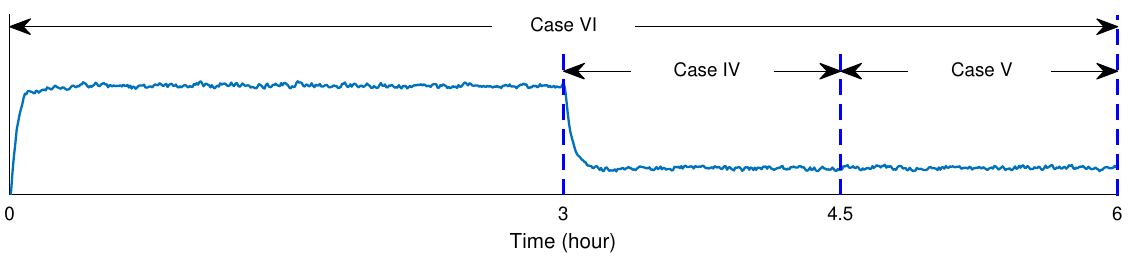}
    }
    \caption{A graphical illustration of the 6 RMSEs for different operating periods under the two case scenarios.} %(The RMSEs are calculated based on 80 hours of model simulations)
    \label{paper1:fig:RMSE:case}
\end{figure*}

\begin{table*}[t!]
  \centering%\small
  \caption{Control performance comparison in terms of RMSEs under different cases and operating periods (each of the RMSE values that are highlighted in bold represents the best result among the three controllers under the same case).}    %Case 1 denotes the control problem with a constant set-point; Case 2 denotes control problem with two set-points.
  \label{paper1:table:rmse}
  \renewcommand\arraystretch{1.5}
    %% local redefinitions
    \renewrobustcmd{\bfseries}{\fontseries{b}\selectfont}
    \renewrobustcmd{\boldmath}{}
    \let\fs\fontsize
    \let\sf\selectfont

    \begin{tabular}{|c|m{1.6cm}<{\centering}|m{1.6cm}<{\centering}|m{1.5cm}<{\centering}|m{1.5cm}<{\centering}|m{1.5cm}<{\centering}|m{1.5cm}<{\centering}|}
    \hline
    %\multirow{3}{*}{Models}
    %  & \multicolumn{3}{c|}{Single set-point} & \multicolumn{3}{c|}{Two set-points}  \\ \cline{2-7}

    %\rule{0pt}{18pt} &\shortstack{Case 1\\ \fs{7pt}{\baselineskip}\sf{(0 - 1.5~hr)}} & \shortstack{Case 2\\\fs{7pt}{\baselineskip}\sf{(1.5 - 80~hr)}} & \shortstack{Case 3\\ \fs{7pt}{\baselineskip}\sf{(0 - 80~hr)}} & \shortstack{Case 4\\ \fs{7pt}{\baselineskip}\sf{(10 - 11.5~hr)}} & \shortstack{Case 5\\\fs{7pt}{\baselineskip}\sf{(11.5 - 80~hr)}} & \shortstack{Case 6\\\fs{7pt}{\baselineskip}\sf{(0 - 80~hr)}} \\ \hline

     & Case I & Case II & Case III & Case IV & Case V & Case VI \\ \hline
%   % This is the 80 hours RMSE results
    %Open-loop & 0.3743 & 0.0196 & 0.05482 & 0.2632 & 0.0193 & 0.06434\\ \hline
%    Full-order KMPC   & \bfseries 0.2439  & 0.0277  & 0.04030 & 0.2512& 0.1026 & 0.10612 \\ \hline
%    %Full-order RKMPC  & 0.2438  & 0.0230  & 0.04045 & 0.2749& 0.1227 & 0.12421 \\ \hline
%    Reduced-order KMPC  & 0.2567  & 0.0146  & 0.03800 & 0.2156& 0.0775 & 0.08459\\ \hline
%    Reduced-order RKMPC & 0.2578  & \bfseries 0.0141 & \bfseries 0.03797 & \bfseries 0.2091&  \bfseries 0.0549 & \bfseries 0.06745\\ \hline

%   % This is the 4/6 hours RMSE results
    %Open-loop           & 0.3742            & 0.0227           & 0.2298           & 0.2666           & 0.0154            & 0.2260           \\ \hline
    Full-order KMPC     & \bfseries 0.2520  & 0.0189           & \bfseries 0.1550 & 0.2436           & 0.1012            & 0.1786           \\ \hline
    Reduced-order KMPC  & 0.2631            & 0.0141           & 0.1615           & 0.2199           & 0.0801            & 0.1718           \\ \hline
    Reduced-order RKMPC & 0.2639            & \bfseries 0.0135 & 0.1620           & \bfseries 0.2134 &  \bfseries 0.0565 & \bfseries 0.1676 \\ \hline
    \end{tabular}
\end{table*}\normalsize

To quantitatively assess and compare the control performance, six operating periods under the two case scenarios (with one single set-point, and with a change in the set-point, respectively) are taken into account. We call them Cases I to VI, which are illustrated in Figure~\ref{paper1:fig:RMSE:case}. Case I, II, and III are defined for a scenario with one single set-point: specifically, Case I considers the first 1.5-hour operation which includes the transient period; Case II considers a 2.5-hour steady-state operation; Case III covers the entire 4-hour operation period. Case IV, V, and VI are defined for another operating condition when there is a change in the set-point: Case IV covers a 1.5-hour period during which the process operation is steered towards the second set-point; Case V considers a 1.5-hour steady-state operation corresponding to the second set-point; Case VI considers the entire 6-hour operation.
% and Case IV to VI are defined under the other scenario as shown in Figure~\ref{paper1:fig:RMSE:case:2sp}. Specifically, Case I is defined on transient-period, which is the first and a half hours; Case II is defined from the first and a half hours until the end to observe the steady-state response; Case III is defined on the entire operating period; Case IV is defined on the transient-period between the two set-points (i.e., from 3th to 4.5th hours); Case V is defined on the subsequent steady-state period after the the change in set-point (i.e., from 4.5th hour to the end); Case VI is defined on the entire operating period. Accordingly, we compute the root mean squared error (RMSE) for each of the six cases.
We compute the root mean squared errors (RMSE) for each of the controllers under the six cases to evaluate the tracking performance. The RMSE is defined as $\text{RMSE}=\sqrt{\frac{1}{nK} \sum_{k=1}^{K}||x_s - x_k ||_2^2}$. Note that since the different magnitudes of the process states, the RMSEs are computed based on scaled system state values for a fair comparison. %In addition, the outcomes were derived from 80 hours of model simulations for both case scenarios. In the scenario with two set-points, the set-point was changed at the tenth hour.

%\textcolor[rgb]{0.00,0.07,1.00}{Table~\ref{paper1:table:rmse} compares the control performance of the three control designs under the two different cases for different operating periods based on root mean squared error (RMSE) . The RMSE is defined as $\text{RMSE}=\sqrt{\frac{1}{nK} \sum_{k=1}^{K}||x_s - x_k ||_2^2}$. Note that the states are scaled for better comparison before calculating the RMSE. Figure~\ref{paper1:fig:RMSE:case} presents six different RMSEs that calculated based on the two different cases and different operating periods. In the one constant set-point case (as shown in Figure~\ref{paper1:fig:RMSE:case:1sp}), RMSE 1 is defined on the first and a half hours to observe the transient-period error; RMSE 2 is defined from the first and a half hours until the end to observe the steady-state response; RMSE 3 is defined on the entire process. When considering the case with two set-points (as shown in Figure~\ref{paper1:fig:RMSE:case:2sp}), emphasis is placed on the performance of the transient-period between the two set-points and the subsequent steady-state period. Subsequently, RMSE 4 is defined on the first and a half hours after changing the set-point (i.e., from 10 to 11.5 hours); RMSE 5 is defined on the steady-state response after the set-point transition (i.e., from 11.5 to 80 hours); RMSE 6 is defined on the whole process. Note that the outcomes were derived from 80 hours of model simulations for both cases. In the case of two set-points, the set-point was changed at the tenth hour.}

The RMSE results for the three controllers under the six considered cases are presented in Table~\ref{paper1:table:rmse}. The proposed reduced-order RKMPC outperforms the other two control methods in four cases out of the six cases. As compared to the full-order KMPC, the proposed reduced-order RKMPC exhibits comparable performance under the scenario with one single set-point and provides better set-point tracking performance under the scenario with a change in set-point. Additionally, the performance of the proposed reduced-order RKMPC is (slightly) better than reduced-order KMPC under Cases II and V.
%, implying that the robust MPC can reduce the approximation errors and bring the closed-loop states closer to the reference values.}
%Table~~\ref{paper1:table:rmse} shows that the reduced-order RKMPC provides slightly better overall and steady-state performance as compared to the reduced-order KMPC and full-order KMPC.
%Furthermore, both the Koopman control models and the proposed reduced-order Koopman control models exhibit comparable performance. % whose errors are lower than the open-loop results.
%It should be mentioned that the RMSE calculated on transient-period is significantly improved, indicating that the control models can facilitate a faster transition from the initial point to the set-point. Especially, for the two set-points case, the performance of the reduced-order RKMPC is much better compared to the reduced-order KMPC and full-order KMPC, implying that the robust MPC can reduce the approximation errors between the reduced-order Koopman linear model and actual nonlinear dynamics.

\subsection{Computation time comparison}\label{paper1:section:cost}
Additionally, the Computation time for the three designs are compared. To ensure a fair comparison of the results, we compute the time costs of the transient-period (first 1.5 hours) with the constant set-point in the simulations. The computation time for one step of each control model is measured through multiple experiments. The simulations are conducted on a PC with an Intel$^{\circledR}$ Core$^{\text{TM}}$ i7-12700 CPU of 12 cores.

\begin{table}[tb!]
  \renewcommand\arraystretch{1.5}
  \centering%\small
  \caption{Comparison of computation time.}\label{paper1:table:speed}
    \begin{tabular}{|c |c|}
      \hline
      Control method & Computation time (s/step)\\ \hline
      Full-order KMPC & 0.0734 \\ \hline
      % RKMPC & 12.4335 \\ \hline
      Reduced-order KMPC & 0.0405  \\  \hline
      Reduced-order RKMPC & 0.0407 \\ \hline
    \end{tabular}
\end{table}

Table \ref{paper1:table:speed} presents the average computation time needed for one-time execution of each of the three control methods. Owing to the reduction in the dimensionality of the model orders as compared to the full-order Koopman model, two reduced-order Koopman-based control models have shorter computation times as compared to the the full-order Koopman-based control. Specifically, the computation times of the reduced-order KMPC and RKMPC are $44.82\%$, $44.55\%$ faster than the full-order KMPC, respectively. % model in terms of computational efficiency, achieving significantly faster computation times.

\section{Conclusion}

In this paper, we proposed a reduced-order Koopman-based robust predictive control approach. The Kalman-GSINDy algorithm was exploited to identify appropriate lifting functions for Koopman modeling. A POD-based reduced-order Koopman modeling approach was proposed, which can be used to construct accurate linear representations of general nonlinear processes while reducing the dimensionality of the resulting linear model compared to the traditional Koopman modeling. A reduced-order robust Koopman MPC design was formulated. A benchmark chemical process was introduced to illustrate the efficacy of the proposed framework. We also compared the performance of the proposed approach with two baselines in terms of tracking performance and computational efficiency. The proposed method can significantly reduce the computational time while providing comparable set-point tracking accuracy.

\section*{Acknowledgment}
This research is supported by Ministry of Education, Singapore, under its Academic Research Fund Tier 1 (RS15/21).

%\bibliographystyle{Unsrt}
%\bibliography{Refv4}

\end{document}